\documentclass{ptephy}
\preprintnumber{XXXX-XXXX}

\usepackage{boites}
\numberwithin{equation}{section}
\usepackage{amsmath}	

\usepackage{amsmath}
\usepackage{amssymb}
\usepackage[subrefformat=parens]{subcaption}

\makeatletter
\DeclareRobustCommand{\cev}[1]{%
  \mathpalette\do@cev{#1}%
}
\newcommand{\do@cev}[2]{%
  \fix@cev{#1}{+}%
  \reflectbox{$\m@th#1\vec{\reflectbox{$\fix@cev{#1}{-}\m@th#1#2\fix@cev{#1}{+}$}}$}%
  \fix@cev{#1}{-}%
}
\newcommand{\fix@cev}[2]{%
  \ifx#1\displaystyle
    \mkern#23mu
  \else
    \ifx#1\textstyle
      \mkern#23mu
    \else
      \ifx#1\scriptstyle
        \mkern#22mu
      \else
        \mkern#22mu
      \fi
    \fi
  \fi
}

\usepackage{version}	
\begin{document}
\title{
Adiabatic approach to large-amplitude collective motion with the
higher-order collective-coordinate operator
}

\author{\name{\fname{Koichi} \surname{Sato}}{1} 
}

\address{\affil{1}{Department of Physics, Osaka City University, Osaka 558-8585, Japan}
\email{satok@sci.osaka-cu.ac.jp}}

\begin{abstract}
We propose a new set of equations to determine the collective Hamiltonian
including the second-order collective-coordinate operator on the basis of the adiabatic
self-consistent collective-coordinate (ASCC) theory.
We illustrate, with the two-level Lipkin model,
that the collective operators including the second-order one 
are self-consistently determined.
We compare the results of the calculations with and without the second-order 
operator and show that, 
without the second-order operator,
the agreement with the exact solution becomes worse as the excitation energy increases,
but that, with the second-order operator included,
the exact solution is well reproduced even for highly excited states.
We also reconsider which equations one should adopt as the basic equations 
in the case where only the first-order operator is taken into account,
and suggest an alternative set of fundamental equations 
instead of the conventional ASCC equations. 
Moreover, we briefly discuss the gauge symmetry of the new basic equations
we propose in this paper.
\end{abstract}

\subjectindex{xxxx, xxx}

\maketitle

\section{Introduction}
In recent papers, we elucidated the relation among
the higher-order collective operators, the $a^\dagger a$ terms, and the gauge symmetry and its breaking
in the adiabatic self-consistent collective-coordinate (ASCC) theory \cite{Sato2015,Sato2017a,Sato2017b}.
The ASCC method~\cite{Matsuo2000} is a practical method of describing the
large-amplitude collective motion,
which is an adiabatic approximation to the SCC method~\cite{Marumori1980} and can be regarded as an advanced version of
the adiabatic time-dependent Hartree-Fock(-Bogoliubov)[ATDHF(B)] theory.

Although several versions of the ATDHF theory had been proposed so far,
they encountered difficulties such as non-uniqueness of the solution 
(See Refs. \cite{Klein1991,Matsuyanagi2010, Nakatsukasa2016} for a review).
As for the ATDHFB theory, which is an extension of the ATDHF theory including the pairing
correlation,
Dobaczewski and Skalski attempted to develop an 
ATDHFB theory assuming the axial quadrupole deformation parameter $\beta$ 
as a collective coordinate \cite{Dobaczewski1981}.
Recently, Li et al also attempted to construct the five-dimensional collective Hamiltonian
on the basis of the ATDHFB theory \cite{Li2012}.
However, the extension of the ATDHF theory to the ATDHFB theory is  not straightforward,
because one needs to decouple the pair-rotational mode from the collective mode of interest. 

In the ASCC method, Matsuo et al \cite{Matsuo2000} first assumed
the gauge angle dependence of the state vector in the form
\begin{align}
|\phi(q,p,\varphi,n)\rangle = e^{-i\varphi\hat N}e^{i\hat G(q,p,n)}|\phi(q)\rangle.
\label{eq:state vector}
\end{align}
Here, $(q,p)$ are collective coordinate and conjugate momentum.
$n=N-N_0$ is the particle number measured from a reference value $N_0$,
and $\varphi$ is the gauge angle conjugate to $n$.
With this form of the state vector,
the collective Hamiltonian is independent of the gauge angle $\varphi$,
and it is guaranteed that the expectation value of the particle number 
is conserved.
Furthermore, there appears no gauge-angle degree of freedom in the equations of motion.
In Ref. \cite{Matsuo2000}, Matsuo et al considered the expansion of $\hat G(q,p,n)$ up to the first order
$\hat G(q,p,n)=p\hat Q(q)+n\hat \Theta(q)$ and 
derived the moving-frame HFB (Hartree--Fock--Bogoliubov) \& QRPA
(quasiparticle random phase approximation) equations, 
which are equations of motion in the ASCC theory.

On the basis of Ref. \cite{Matsuo2000}, Hinohara et al performed a numerical calculation
and encountered a difficulty of finding the solution due to the numerical instability~\cite{Hinohara2007}.
They found that this instability was caused by the symmetry 
under the following transformation under which the basic equations are invariant if 
the collective-coordinate and particle number operators commute, i. e. $[\hat Q,\hat N]=0$.
\begin{align}
&\hat{Q} \rightarrow \hat{Q} +\alpha \hat{N} \label{eq:^Q' ex1},\\
&\hat{\Theta} \rightarrow \hat{\Theta} +\alpha \hat{P} \label{eq:^Theta' ex1},\\
&\lambda \rightarrow \lambda -\alpha \partial_q V, \\
&\partial_q \lambda \rightarrow \partial_q\lambda -\alpha C
 \label{eq:dqlambda' ex1}
\end{align}
As this transformation changes the phase of the state vector,
they called it the ''gauge'' transformation and proposed a gauge-fixing
prescription to remove the redundancy associated with the gauge symmetry.
Their prescription for the gauge fixing is as follows.
While the moving-frame HFB \& QRPA equations are invariant under the above
transformation at the HF equilibrium point $\partial_q V=0$, 
it is not the case at non-equilibrium points unless $[\hat Q,\hat N]=0$.
Therefore, they first require the commutativity of the
collective-coordinate and particle-number operators $[\hat Q,\hat N]=0$ for the gauge symmetry
of the moving-frame HFB \& QRPA equations, and then fix the gauge.
With this prescription, they succeeded in obtaining the solution and 
applied the one-dimensional~(1D) ASCC method to the multi-$O(4)$ model
\cite{Hinohara2007} and the oblate-prolate shape-coexistence phenomena
in the proton-rich Se and Kr isotopes~\cite{Hinohara2008,Hinohara2009}
(Here, we mean by the $D-$dimensional ASCC method that the dimension of the
collective coordinate $q$ is $D$.)

After the successful application of the 1D ASCC method, an
approximate version of the 2D ASCC method, the constrained
HFB plus local QRPA method, was proposed and applied to large-amplitude
quadrupole collective
dynamics~\cite{Hinohara2010b,Sato2011,Watanabe2011,Hinohara2011a,Hinohara2011,Hinohara2012,Yoshida2011,
Sato2012}.
Moreover, the 1D ASCC method without the pairing correlation 
was also applied to nuclear reaction~\cite{Wen2016,Wen2017}
(In these calculations, the so-called curvature term was neglected.)
However, little progress had been made in understanding of the gauge symmetry,
until very recently  
we analyzed the gauge symmetry and its breaking under more general gauge
transformations in the ASCC theory,
on the basis of the Dirac-Bergmann theory of the constrained systems~\cite{Sato2015,Sato2017a}.
There, it was shown that the gauge symmetry in the ASCC method is
broken by the adiabatic approximation, and that the gauge symmetry is
partially retained by containing the higher-order collective operators
in the adiabatic expansion.

According to the generalized Thouless theorem
(Refs. \cite{Thouless1960,Marumori1980,Rowe1980,
Ring1977, Suzuki1983}),
$\hat G$ in Eq. (\ref{eq:state vector}) can be written in terms of only
$a^\dagger a^\dagger$ 
and $aa$ terms.
We shall call $a^\dagger a^\dagger$ and $aa$ terms A-terms and 
$a^\dagger a$ and $aa^\dagger$ (or equivalently $a^\dagger a$ and constant terms)
B-terms, respectively.
In Hinohara's prescription, the commutativity $[\hat Q,\hat N]=0$ is required, 
which implies that one needs to introduce B-terms  in $\hat Q$ in
contrast with the theorem.
Thus, there are two approaches to conserve the gauge symmetry:
one is the approach with higher-order operators consisting of only
A-terms (Approach A),
and the other with only the first-order operator containing B-terms as
well as A-terms (Approach B). The relation between the two approaches
was investigated in Ref. \cite{Sato2017b}, and  it was shown that the inclusion of the B-terms in
the collective-coordinate operator is equivalent to that of a certain
kind of the higher-order operators. 

What should be emphasized is that these higher-order collective operators and
B-terms contribute to the equations of motion and the inertial mass directly.
As shown in Ref. \cite{Sato2015}, the gauge symmetry
is associated with the constraint on the particle number and appears
when the pairing correlation is taken into account.
However, the higher-order operators and B-terms can affect dynamics
through the equations of motion and the inertial mass
even if there is no pairing correlation. 
In fact, in the ATDHF theories proposed in the 1970's, similar things
had been recognized.
In his paper on the ATDHF in 1977 (Ref. \cite{Villars1977}),
Villars mentioned the extension of his ATDHF theory 
including the higher-order operator 
(more strictly, the extension with the first- and third-order operators
and no second-order operator) and preannounced a publication on it
: ``Ref. 17) A. Toukan and F. Villars, to be published'' in Ref. \cite{Villars1977}. 
However, as far as the author knows, it was not published after all.
It is noteworthy that the second-order operator was not included in his extension.
In this paper, we include the second-order operator because it does
contribute to the equations of motion and the inertial mass.
On the other hand, in the ATDHF theory by Baranger and V\'en\'eroni (Ref. \cite{Baranger1978}),
they proposed the density matrix in the form of
$\rho(t)=e^{i\chi(t)}\rho_0 (t) e^{-i\chi(t)}$ with Hermitian and time-even 
$\rho_0(t)$ and $\chi(t)$.
They emphasized that $\chi(t)$ can be written in terms of A-terms only,
but included B-terms as well as A-terms in the treatment 
of the translational motion.
Thus,  although the necessity of the inclusion of the higher-order
operators or B-terms was already recognized in the ATDHF theories in the
1970's, there has been no general theory to determine them.  

Now that the relation between the higher-order operators and B-terms has
been revealed in Ref. \cite{Sato2017b},  
we propose in this paper a new set of basic equations to determine the higher-order
operator, which is different from Hinohara's prescription.
In Ref. \cite{Sato2017b}, it is shown that,
with Hinohara's prescription,  
one can take into account the contribution from the second- and
third-order collective operators 
in an effective way by including the B-part of the collective-coordinate operator $\hat Q_B$.
That is an advantage of Hinohara's prescription.
In the case without the pairing correlation, however, 
one cannot determine the higher-order collective operators
through Hinohara's prescription
because $[\hat Q,\hat N]=0$ is  automatically satisfied without $\hat Q_B$.
(In the first place, there exists no gauge symmetry when the pairing
correlation is not taken into account, and there is no reason to require
$[\hat Q,\hat N]=0$.)
Without the pairing correlation, no way has been known to include the
higher-order contribution. 
In this sense, the cases with and without the pairing correlation
have not been treated on an equal basis so far.
%
Hinohara's prescription is based on Approach B, in which the B-part of
the collective-coordinate operator $\hat Q_B$ is introduced.
It is also worth mentioning that the Hilbert space in Approach A is larger than
that in Approach B because only the higher-order collective operator
which are written in terms of (multiple) commutators of $\hat Q_A$ and
$\hat Q_B$ can be taken into account in Approach B.
In this paper, we employ Approach A.

In the conventional ASCC theory without the pairing correlation, 
only the first-order operators consisting of A-terms are taken into
account, and the contributions from the higher-order operators to 
the equations of motion and the inertial mass are missing.
However, it cannot be neglected from a simple order counting
in the adiabatic expansion. 
In Refs. \cite{Wen2016,Wen2017}, Wen and Nakatsukasa 
successfully reproduced the inertial mass against the translational motion
without including the higher-order collective operators.
Actually, for the mode with $\partial_q V=0$, the higher-order operators
and B-terms do not contribute to the inertial mass.
The reason for this is explained in Ref. \cite{Sato2017b} 
from the viewpoint of Approach B, 
but it also applies to Approach A straightforwardly.
For general collective modes of interest, however,
$\partial_q V=0$ is not satisfied, and a theory with which one can correctly 
evaluate the contribution from the higher-order operators is necessary.
In Ref. \cite{Wen2017},  Wen and Nakatsukasa failed to find the solution 
at a certain point on the collective path and pointed out a possibility
to solve this problem by including the pairing correlation.
Depending on the particle number, or on the collective
coordinate even in one nucleus, the pairing gap changes and can vanish.
Therefore, a theory is favorable with which one can treat the cases with and without
the pairing correlation on an equal footing. 
In this paper, we first consider the case without the pairing correlation
on the basis of Approach A including the second-order
collective-coordinate operator $\hat Q^{(2)}$ 
and propose a new set of fundamental equations to determine the
collective operators. The set of fundamental equations for the case
with the pairing correlation included is derived in a straightforward
way.

The paper is organized as follows.
Sect. 2 describes the formulation.
In Sect. 2.1, we propose a set of the basic equations including
the second-order collective-coordinate operator $\hat Q^{(2)}$. 
In Sect. 2.3, we introduce the two-level Lipkin model,  
and it is shown in Sect. 2.3 that the basic equations proposed in Sect. 2.1 are reduced to one
differential equation in the case of the two-level Lipkin model.
We give the collective Schr\"odinger equation in Sect. 2.4, 
and the solution to the conventional ASCC equations without $\hat
Q^{(2)}$ is given in Sect. 2.5.
The numerical results are shown in Sect. 3.
We compare the calculations with and without $\hat Q^{(2)}$ 
employing the Lipkin model. The numerical results show that, 
for low-energy states, both of the calculations reproduce the exact
solution well, but that, with increasing the excitation energy,
the agreement with the exact solution becomes worse when the
second-order collective operator $\hat Q^{(2)}$ is neglected.
With $\hat Q^{(2)}$, the exact solution is well reproduced even for
higher excited states.
In Sect. 4, we consider which equations should be adopted as 
the basic equations of motion when only the first-order operator
$\hat Q^{(1)}$ is included, taking the Lipkin model as a simple example.
We propose an alternative set of the basic equations including
an equation introduced in Sect. 2 
instead of the conventional moving-frame RPA equation of $O(p^2)$.
In Sect. 5, we briefly discuss the gauge symmetry of the basic equations 
derived in Sect. 2 with the pairing correlation included.
Concluding remarks are given in Sect. 6.
In Appendix A, the expressions of the derivatives of the collective operators
are given in terms of quasispin operators in the Lipkin model. In
Appendix B, the potential curvature $C$ in the $q$ space with $B(q)=1$ is given
in the case where only the first-order operator is taken into account.


\section{Formulation}
\subsection{Basic equations with the second-order operator $\hat Q^{(2)}$}

First, we consider the equations of motion without pairing correlation.
In preceding papers~\cite{Sato2017a,Sato2017b},
we considered the ASCC theory with the second- and third-order
collective-coordinate operators.
Here we adopt the approach with the higher-order operators consisting
of only A-terms, i. e., Approach A.
The state vector in the ASCC theory is given by
\begin{align}
|\phi (q,p) \rangle =e^{i\hat G(q,p)}|\phi (q) \rangle.
\label{eq:state vector without pairing}
\end{align}
Here, $\hat G$ is expanded as
\begin{align}
 \hat G(q,p) 
=p \hat Q^{(1)}(q)
+\frac{1}{2}p^2 \hat Q^{(2)}(q) 
+\frac{1}{3!}p^3 \hat Q^{(3)}(q), 
\end{align}
with
\begin{align}
 \hat Q^{(i)}(q)
=\sum_{\alpha\beta} Q^{(i)}_{\alpha\beta}a_\alpha^\dagger
 a_\beta^\dagger
+Q^{(i)*}_{\alpha\beta}a_\beta a_\alpha \,\,\,(i=1,2,3).
\end{align}
The equations of motion in the ASCC theory is derived from 
the invariance principle of the Schr\"odinger equation
\begin{align}
 \delta\langle \phi(q,p)|i\partial_t -\hat H 
|\phi(q,p)\rangle = 0,
\label{eq:TDVP}
\end{align}
which can be rewritten into the equation of collective submanifold (CS):
\begin{align}
  \delta \langle \phi(q,p)|\hat H -
  \frac{\partial \mathcal{H}}{\partial p}\mathring{P}
- \frac{\partial \mathcal{H}}{\partial q}\mathring{Q}
|\phi(q,p) \rangle  =0, \label{eq:coll.man2} 
\end{align}
with $(\mathring{P}, \mathring{Q}):=(i\partial_q, -i\partial_p)$.
The collective Hamiltonian is given by
\begin{align}
\mathcal{H}(q,p) &=  V(q)
+\frac{1}{2}B(q)p^2 
\end{align}
with
\begin{align}
 V(q)&=\langle\phi (q)|\hat H |\phi(q)\rangle ,\\
 B(q)&=\langle\phi (q)|[\hat H, i\hat  Q^{(2)}] |\phi(q)\rangle 
      -\langle\phi (q)|[[\hat H, \hat  Q^{(1)}],\hat Q^{(1)}]
 |\phi(q)\rangle. \label{eq:B(q)} 
\end{align}
One can see that the second-order operator $\hat Q^{(2)}$ contributes to the inertial mass,
and that one cannot neglect the term involving $\hat Q^{(2)}$ in $B(q)$
from a simple order counting,
as both of the first and second terms of the right-hand side of Eq. (\ref{eq:B(q)})
are $O(p^2)$ terms.

By substituting the state vector (\ref{eq:state vector without pairing}) 
into Eq. (\ref{eq:coll.man2}) and expanding in powers of $p$,
the moving-frame HF \& RPA equations, which are
the equations of motion in the ASCC theory (without the pairing),
 are obtained as

\noindent
\underline{Moving-frame HF equation}
\begin{equation}
 \delta \langle \phi(q)|\hat H -\partial_q V\hat Q^{(1)}
  |\phi(q)\rangle =0,
\label{eq:moving-frame HF}
\end{equation}

\noindent
\underline{Moving-frame RPA equations}
\begin{equation}
 \delta \langle \phi(q)|[\hat H , \hat Q^{(1)}] 
-\frac{1}{i}B(q)\hat P -\frac{1}{i}\partial_q V \hat Q^{(2)} |\phi(q)\rangle =0,
\label{eq:moving-frame RPA1 with Q2}
\end{equation}

\begin{align}
& \delta \langle \phi(q)|
[\hat H -\partial_q V \hat Q^{(1)}, \frac{1}{i}B\hat P]
-B(q)C(q)\hat Q^{(1)}\notag\\
&\hspace{6em}-\frac{1}{2}
\partial_q V
\left\{[[\hat H,  \hat Q^{(1)}],\hat Q^{(1)}] + [\hat H, \frac{1}{i}\hat
 Q^{(2)}] \right.\notag\\
&\hspace{12em}\left.+\partial_qV\left(
\hat Q^{(3)}+\frac{1}{2} [\hat Q^{(1)}, \frac{1}{i}\hat Q^{(2)}]\right)\right\}
|\phi(q)\rangle =0, \label{eq:moving-frame RPA2 with Q2}
\end{align}
where
\begin{align}
 C:=\partial_q^2V(q)-\Gamma(q)\partial_qV:=\partial_q^2V(q)+\frac{1}{2B}\partial_qB\partial_qV.
\label{eq:definition of C}
\end{align}

The moving-frame HF \& RPA equations are derived from the $O(1), O(p)$ and
$O(p^2)$ expansions of the equation of CS (\ref{eq:coll.man2}).
While the moving-frame HF equation (\ref{eq:moving-frame HF}) is the zeroth-order 
equation of CS  with respect to $p$, i. e., the $O(1)$ terms of Eq. (\ref{eq:coll.man2}),
the moving-frame RPA equation of $O(p)$ (\ref{eq:moving-frame RPA1 with
Q2}) 
is the first-order equation of CS.
On the other hand, the moving-frame RPA equation of $O(p^2)$
(\ref{eq:moving-frame RPA2 with Q2})
is not the $O(p^2)$ expansion of the equation of CS (\ref{eq:coll.man2})
itself, 
which is given by Eq. (\ref{eq:p^2 EoCSM}),
but is derived from
the second-order equation of CS (\ref{eq:p^2 EoCSM})
and the $q$-derivative of the zeroth-order equation of CS, i. e., 
the moving-frame HF equation (\ref{eq:moving-frame HF}).
(See Ref. \cite{Sato2017a} for the detailed derivation.)
The fourth- and higher-order operators $\hat Q^{(i)}(i\geq 4)$ do not contribute
to the equations of motion up to this order.


Here we expand $\hat G$ up to $\hat Q^{(2)}$ and omit $\hat
Q^{(i)}(i\geq 3)$.
$\hat Q^{(3)}$ contributes to the moving-frame RPA equation of $O(p^2)$
(\ref{eq:moving-frame RPA2 with Q2}) but does not 
contribute to the equations of motion of 
$O(1)$ and $O(p)$ and the inertial mass.
When the expansion up to $\hat Q^{(2)}$ is taken,
there are one more unknown operators than in the conventional ASCC
equations taking up to $\hat Q^{(1)}$.
Thus, one more equation is necessary to determine 
the collective operators and the state vector self-consistently.
The moving-frame equations are derived by expanding the equation of 
CS (\ref{eq:coll.man2}) in powers of $p$. 
The $O(p^3)$ equation of CS is
a possible candidate for such an equation to add to the three moving-frame 
equations.
However, as easily confirmed, there appears $\hat Q^{(4)}$ as well as
$\hat Q^{(3)}$ in the $O(p^3)$ equation of CS,
so one needs to make an approximation neglecting 
$\hat Q^{(3)}$ and $\hat Q^{(4)}$ to close the set of the equations.

Therefore, we adopt another equation, that is, the $q$-derivative
of the moving-frame HF equation.
As mentioned above,
the moving-frame RPA equations of $O(p^2)$
(\ref{eq:moving-frame RPA2 with Q2})
is derived from the second-order equation of CS and
the $q$-derivative of the moving-frame HF equation \cite{Matsuo2000,Sato2017a}.
Here we consider the following equations as a set of the basic equations in the case
where $\hat Q^{(2)}$ is included. 
\begin{equation}
 \delta \langle \phi(q)|\hat H -\partial_q V\hat Q^{(1)}
  |\phi(q)\rangle =0,
\label{eq:mfHF}
\end{equation}
\begin{equation}
 \delta \langle \phi(q)|[\hat H , \hat Q^{(1)}] 
-\frac{1}{i}B(q)\hat P -\frac{1}{i}\partial_q V \hat Q^{(2)} |\phi(q)\rangle =0,
\label{eq:mfRPA1 with Q2}
\end{equation}
\begin{align}
 \delta \langle \phi(q)|
[\hat H -\partial_q V \hat Q^{(1)}, \frac{1}{i}B(q)\hat P]
-B(q)C(q)\hat Q^{(1)}-\partial_q V B(q)D_q\hat Q^{(1)}
|\phi(q)\rangle =0,
\label{eq:dq moving-frame HF eq.}
\end{align}
\begin{align}
 \delta \langle \phi(q)|\frac{1}{2}[[\hat H,\hat Q^{(1)}], \hat Q^{(1)}] 
-B(q) D_q\hat Q^{(1)} 
-\frac{i}{2}[\hat H ,\hat Q^{(2)}]  
-\frac{i}{4}\partial_q V[\hat Q^{(1)} ,\hat Q^{(2)}]  
|\phi(q)\rangle =0,  \label{eq:p^2 EoCSM}  
\end{align}
where the covariant derivative of $\hat Q^{(1)}$ is given by
\begin{align}
D_q \hat Q^{(1)}= \partial_q \hat Q^{(1)}+\Gamma (q)\hat Q^{(1)}.
\end{align}
Eq. (\ref{eq:p^2 EoCSM}) is the second-order equation of collective
submanifold with $\hat Q^{(3)}$ omitted.
The term $[\hat Q^{(1)},\hat Q^{(2)}]$ is kept in Eq. (\ref{eq:p^2 EoCSM})
but does not contribute because it is a B-term (See Ref. \cite{Sato2017b}).
The moving-frame RPA equation of $O(p^2)$ (\ref{eq:moving-frame RPA2 with Q2})
is obtained by eliminating $D_q\hat Q^{(1)}$ from 
Eq. (\ref{eq:dq moving-frame HF eq.})
using Eq. (\ref{eq:p^2 EoCSM}) multiplied by $\partial_q V$.
Hence, for $\partial_q V \neq 0$, 
it is equivalent to adopt Eqs. (\ref{eq:mfRPA1 with Q2}) and (\ref{eq:dq
moving-frame HF eq.}) as basic equations
to adopting Eqs. (\ref{eq:mfRPA1 with Q2})
and (\ref{eq:moving-frame RPA2 with Q2}) with $Q^{(3)}$ omitted.
In Ref. \cite{Wen2016,Wen2017} by Wen and Nakatsukasa,
Eq. (\ref{eq:dq moving-frame HF eq.}) without the $D_q\hat Q^{(1)}$ term
is called ``moving'' RPA equation of $O(p^2)$.
It is somewhat misleading, however, because
this equation does not contain $O(p^2)$ terms.
Eq. (\ref{eq:dq moving-frame HF eq.}) is derived from the first derivative
of the equation of CS with respect to $q$, 
and thus it would rather be called the equation of ''$O(q)$''.
The other moving-frame RPA equation (\ref{eq:mfRPA1 with Q2}) is of
$O(p)$. 
At the HF equilibrium point $\partial_q V=0$, they reduce to the ordinary RPA equations,
in which the collective coordinate and momentum operators are involved
in a symmetric form.
At the equilibrium point $\partial_q V=0$,
the moving-frame HF equation (\ref{eq:moving-frame HF}) reduces to the
ordinary HF equation.
Eq. (\ref{eq:p^2 EoCSM}) gives the relation between $\partial_q \hat Q^{(1)}$ and
$\hat Q^{(2)}$.

In the above equations, there appears the covariant derivative of 
$\hat Q^{(1)}$.
In actual calculations, one can choose a coordinate system with $B(q)\equiv 1$
by the scale transformation of $q$.
Here we mean by the symbol $\equiv$ that the inverse inertial mass $B(q)$ is
unity everywhere along the collective path.
Then, the covariant derivative reduces to the ordinary partial
derivative,
and the equations may be solved by approximating $\partial_q\hat Q^{(1)}(q)$ by finite difference. 
The simplest scheme of the finite difference is
$\partial_q\hat Q^{(1)}(q) = \left[\hat Q^{(1)}(q) - \hat Q^{(1)}(q-\delta q)\right]/\delta q$.
At the HF(B) equilibrium point ($\partial_q V=0$),
Eqs. (\ref{eq:mfHF})-(\ref{eq:dq moving-frame HF eq.}) reduces to the HF \& RPA equations, 
and $(\hat Q^{(1)},\hat P)$ can be obtained.
Starting from the equilibrium point and approximating the derivative of
the collective-coordinate operator
$\partial_q \hat Q^{(1)}$ with the finite difference, one can solve the
above set of the equations.

Above we have considered the case without pairing correlation,
the basic equations in the  case with the pairing correlation can be 
obtained in a straightforward way as follows.
\begin{equation}
 \delta \langle \phi(q)|\hat H_M 
  |\phi(q)\rangle =0,
\label{eq:moving-frame HFB}
\end{equation}
\begin{equation}
 \delta \langle \phi(q)|[\hat H_M, \hat Q^{(1)}] 
-\frac{1}{i}B(q)\hat P -\frac{1}{i}\partial_q V \hat Q^{(2)} |\phi(q)\rangle =0,
\label{eq:moving-frame QRPA1 with Q2}
\end{equation}
\begin{align}
 \delta \langle \phi(q)|
[\hat H_M, \frac{1}{i}B(q)\hat P]
-B(q)C(q)\hat Q^{(1)}-\partial_q V B(q)D_q\hat
 Q^{(1)}-\partial_q\lambda \tilde N
|\phi(q)\rangle =0,
\label{eq:dq moving-frame HFB eq.}
\end{align}
\begin{align}
 \delta \langle \phi(q)|\frac{1}{2}[[\hat H_M ,\hat Q^{(1)}], \hat Q^{(1)}] 
-B(q) D_q\hat Q^{(1)} 
-\frac{i}{2}[\hat H-\lambda \tilde N ,\hat Q^{(2)}]  
-\frac{i}{4}\partial_q V[\hat Q^{(1)} ,\hat Q^{(2)}]  
|\phi(q)\rangle =0,  \label{eq:p^2 EoCSM with pairing}  
\end{align}
where the moving-frame Hamiltonian is given by
\begin{align}
 \hat H_M = \hat H -\lambda \tilde N -\partial_q V \hat Q^{(1)},
\end{align}
with $\tilde N$ is the particle-number operator measuring from $N_0$, 
$\tilde N=\hat N-N_0$.
In Sect. 5, we shall give a brief consideration on the gauge symmetry of these equations.


\subsection{The Lipkin model}
In this section, we introduce the Lipkin model with two $N$-fold degenerate
levels~\cite{Lipkin1965}.
We follow the formulation of Ref. \cite{Holzwarth1973} by Holzwarth.
We label the states in the upper and lower level by $p=1,2, ... , N$
and $-p$, respectively. 

In this model, the Hamiltonian is given by
\begin{align}
H&=\frac{1}{2}\epsilon \sum_{p>0}  (c_p^\dagger c_p -c_{-p}^\dagger
 c_{-p})+\frac{1}{2}V\sum_{p,p^\prime>0}(c_p^\dagger c_{p^\prime}^\dagger
 c_{-p^\prime} c_{-p} + c_{-p}^\dagger c_{-p^\prime}^\dagger
 c_{p^\prime} c_{p}) \notag\\
&=\epsilon \hat K_0 + \frac{1}{2}V (\hat K_+ \hat K_+ +\hat K_- \hat K_- ),
\label{eq:H_lipkin}
\end{align}
with
\begin{align}
\hat K_0&=\frac{1}{2}\sum_{p>0}(c_p^\dagger c_p- c_{-p}^\dagger c_{-p}), \\
\hat K_+&=\sum_{p>0} c_p^\dagger c_{-p}, \,\,\, \hat K_-=\hat
 K_+^\dagger .
\end{align}
The quasispin operators satisfy
\begin{align}
 [\hat K_+,\hat K_-]=2\hat K_0,\,\,\, [\hat K_0,\hat K_{\pm }]=\pm \hat K_{\pm}.
\end{align}

We assume that the system contains $N$ particles. 
If there is no interaction $V=0$, the lower levels are fully occupied
and the upper levels are completely empty in the ground state. 
We denote the ground state in the non-interacting case by 
$|0\rangle$. Then, we have
\begin{align}
 c_p|0\rangle =0, \,\,\, c_{-p}^\dagger |0\rangle =0, \,\,\, \hat K_-|0\rangle=0.
\end{align}
We define the particle/hole creation and annihilation operators
$a_{\pm p}$ as follows:
\begin{align}
 a_{p}&=c_{p}, &a_{p}^\dagger=c_p^\dagger ,\\
 a_{-p}&=c_{-p}^\dagger, &a_{-p}^\dagger=c_{-p}.
\end{align}
Since $a_{\pm p}|0\rangle =0$, $|0\rangle$ is the vacuum with respect to
$a_{\pm p}$.
$a_{\pm p}$ satisfies the canonical commutation relations of fermions.
With $a_{\pm p}$, the quasispin operators are written as
\begin{align}
\hat K_0&=\frac{1}{2}\sum_{p>0}(a_p^\dagger a_p- a_{-p}a_{-p}^\dagger)
=\frac{1}{2}\sum_p(a_p^\dagger a_p+ a_{-p}^\dagger a_{-p})-\frac{N}{2},\\
\hat K_+&=\sum_{p>0} a_p^\dagger a_{-p}^\dagger, \,\,\, \hat K_-=\sum_{p>0} a_{-p}a_p .
\end{align}

Following Ref. \cite{Holzwarth1973},  we introduce the ''deformed'' state as
\begin{align}
 |a\rangle = \exp (a \hat K_+)|0\rangle, \label{eq:Ho14}
\end{align}
and the ``deformed'' operators as
\begin{align}
\begin{pmatrix}
 \alpha_{p}^\dagger \\
 \alpha_{-p}
\end{pmatrix} 
=
\begin{pmatrix}
 \cos \alpha && -\sin\alpha e^{-i\psi} \\
 \sin \alpha e^{i\psi} && \cos\alpha
\end{pmatrix} 
\begin{pmatrix}
 a_p^\dagger \\
 a_{-p}
\end{pmatrix}\label{eq:Ho15} .
\end{align}
%
With the deformed operators, 
we shall define
the new quasispin operators:
\begin{align}
\hat I_0&=\frac{1}{2}\sum_{p>0}(\alpha_p^\dagger \alpha_p-
 \alpha_{-p} \alpha_{-p}^\dagger), \\
\hat I+&=\sum_{p>0} \alpha_p^\dagger \alpha_{-p}^\dagger, \,\,\, \hat I_-=\hat
 I_+^\dagger =\sum_{p>0}\alpha_{-p}\alpha_p,
\end{align}
which are connected with the undeformed quasispin operators by
\begin{align}
 \begin{pmatrix}
  \hat I_0\\ \hat I_+ \\ \hat I_- 
 \end{pmatrix}
=
 \begin{pmatrix}
  \cos\phi && -\frac{1}{2}e^{i\psi} \sin\phi &&
  -\frac{1}{2}e^{-i\psi}\sin\phi \\ 
  e^{-i\psi}\sin\phi && \cos^2\frac{1}{2}\phi&& - e^{-2i\psi}\sin^2\frac{1}{2}\phi \\ 
    e^{i\psi}\sin\phi && -e^{2i\psi}\sin^2\frac{1}{2}\phi&&  \cos^2\frac{1}{2}\phi 
 \end{pmatrix}
 \begin{pmatrix}
  \hat K_0\\ \hat K_+ \\ \hat K_- 
 \end{pmatrix}\label{eq:Hol(16)}.
\end{align}
The inverse transformation is given by
\begin{align}
  \begin{pmatrix}
  \hat K_0\\ \hat K_+ \\ \hat K_- 
 \end{pmatrix}
=
\begin{pmatrix}
\cos \phi && \frac{1}{2}e^{i\psi}\sin\phi && \frac{1}{2}e^{-i\psi}\sin \phi \\
-e^{-i\psi}\sin\phi && \cos^2 \frac{1}{2}\phi && -e^{-2i\psi}\sin^2 \frac{1}{2}\phi \\
-e^{-i\psi}\sin \phi && -e^{2i\psi}\sin^2\frac{1}{2}\phi && \cos^2 \frac{1}{2}\phi
\end{pmatrix}
 \begin{pmatrix}
  \hat I_0\\ \hat I_+ \\ \hat I_- 
 \end{pmatrix}.
\end{align}
As shown in Ref. \cite{Holzwarth1973},
$|a\rangle$ is not normalized, and the normalized deformed state is
given by
\begin{align}
 |\phi\rangle = \left(\cos \frac{1}{2}\phi\right)^N|a\rangle .
\end{align}

As in Ref. \cite{Holzwarth1973}, 
we set  $\psi=0$ and denote
the quasispin operators as
$\hat J_i(=\hat I_i|_{\psi=0})$.
Then the Hamiltonian can be written as
\begin{align}
\hat H&= 
2E\hat J_0
+\frac{1}{2}\sin\phi\epsilon\left(1 -\chi\cos\phi\right)(\hat J_+ +\hat J_-)\notag\\
&+\frac{1}{4}V\sin^2\phi \left(4\hat J_0^2 +(4N-2) \hat J_0 \right)
-\frac{1}{2}V\sin^2\phi \hat J_+\hat J_-\notag\\
&-V\sin\phi\cos\phi \left[ \hat J_+\hat J_0 +\hat J_0\hat J_- + \frac{N}{2}  \left(\hat J_+ +\hat J_- \right)\right] 
+\frac{1}{2}V\left(1-\frac{1}{2}\sin^2\phi\right)(\hat J_+\hat J_+ +\hat J_-\hat J_-),
\label{eq:H in terms of hat J0}
\end{align}
where
\begin{align}
2E&=\epsilon\left(\cos\phi+\chi\sin^2\phi\right),\notag\\
\chi&=(1-N)V/\epsilon.
\end{align}
We shall define 
\begin{align}
 \tilde J_0=\hat J_0+\frac{N}{2},
\end{align}
and rewrite the Hamiltonian
as follows.
\begin{align}
\hat H&= V(\phi)+
2E\tilde J_0
+\frac{1}{2}\epsilon\sin\phi\left(1-\chi\cos\phi\right)(\hat J_+ +\hat J_-)\notag\\
&+\frac{1}{4}V\sin^2\phi \left(4\tilde J_0^2 -2 \tilde J_0 \right)
-\frac{1}{2}V\sin^2\phi \hat J_+\hat J_- \notag\\
&-V\sin\phi\cos\phi [ \hat J_+\tilde J_0 +\tilde J_0\hat J_- ]  
+\frac{1}{2}V\left(1-\frac{1}{2}\sin^2\phi\right)(\hat J_+\hat J_+ +\hat J_-\hat J_-).
\label{eq:H in terms of tilde J0}
\end{align}
As easily confirmed, the last four terms in Eq.
(\ref{eq:H in terms of tilde J0})
are normally-ordered quartic terms of $(\alpha, \alpha^\dagger)$.
Here, $V(\phi)$ corresponds to the collective potential,
and is given by
\begin{align}
V(\phi)&=\langle \phi | \hat H | \phi \rangle 
= -\frac{N}{2}\epsilon (\cos\phi +\frac{1}{2}\chi\sin^2\phi ).
\label{eq:HF energy}
\end{align}
The reader should not confuse $V(\phi)$ with the interaction parameter $V$.

\begin{figure}[tbp]
\begin{tabular}{cc}
\begin{minipage}{0.48\hsize}
\centering
\includegraphics[width=\textwidth]{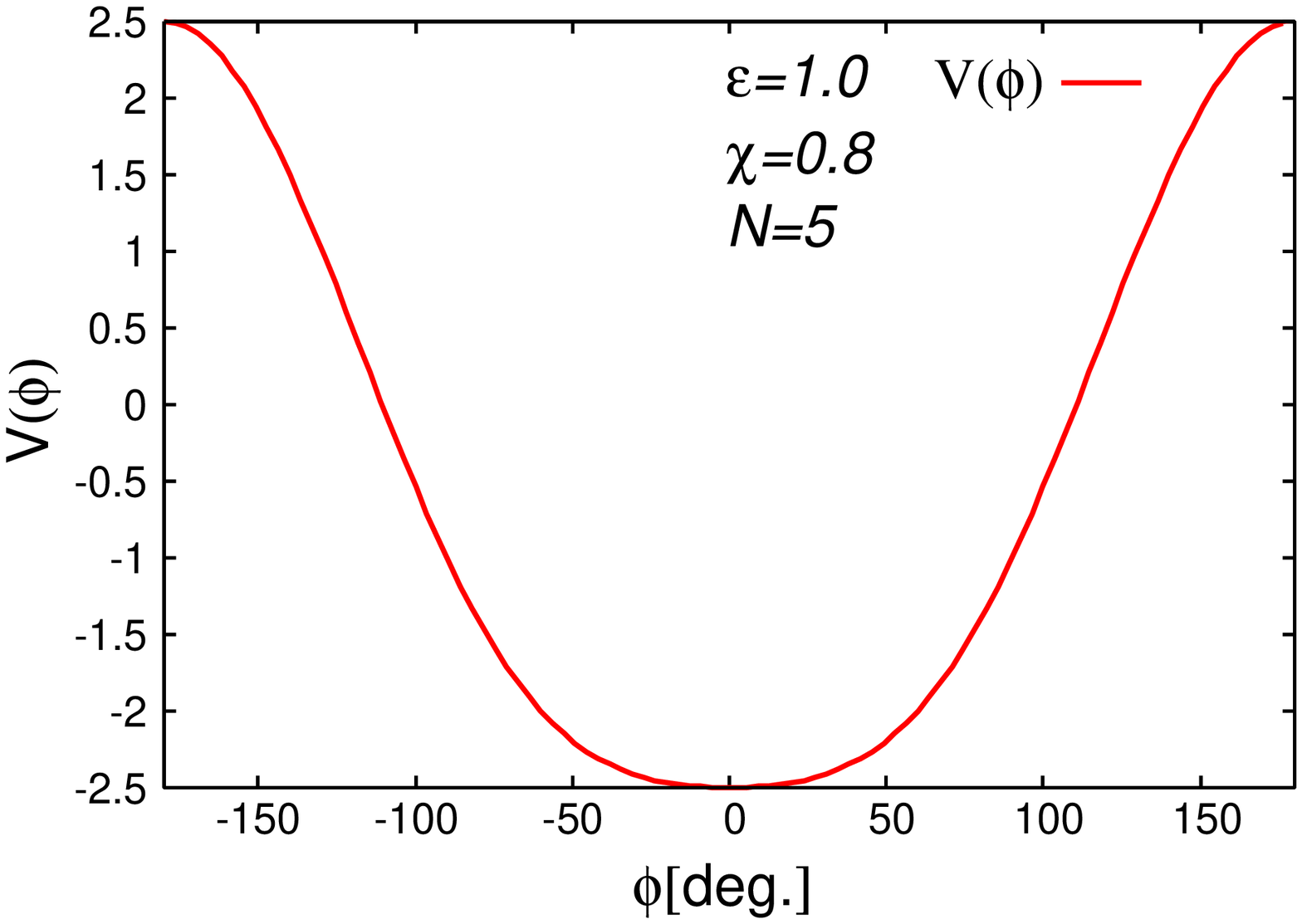}
\caption{$V(\phi)$ with $N=5,\epsilon=1.0,\chi=0.8.$}
\label{fig:N5chi0.8e1_V}
\end{minipage}
\hspace*{0.02\hsize}
\begin{minipage}{0.48\hsize}
\begin{center}
\includegraphics[width=\textwidth]{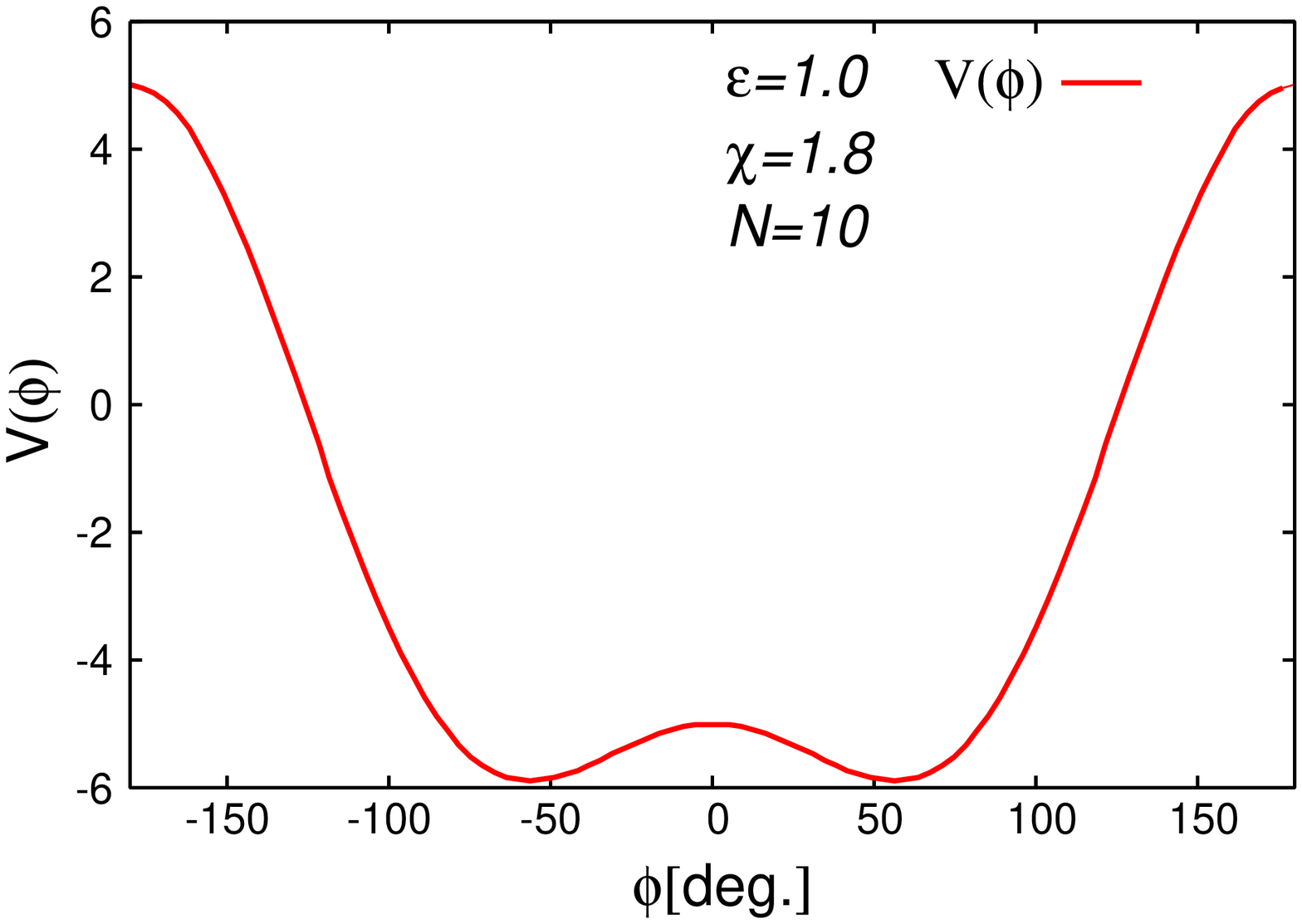}
\end{center}
\caption{$V(\phi)$ with $N=10,\epsilon=1.0,\chi=1.8.$}
\label{fig:N10chi1.8e1_V}
\end{minipage}
\end{tabular}
\end{figure}

The Hartree--Fock equations is given by the condition for $V(\phi)$ to
take an extremum,
\begin{align}
\frac{\partial V}{\partial \phi}= \frac{N}{2}\epsilon\sin\phi\left(1
 -\chi\cos\phi \right)=0,
\label{eq:HF condition}
\end{align}
which is equivalent to the condition for
the third term in Eq. (\ref{eq:H in terms of tilde J0}) to vanish.
Examples of the collective potential 
for $\chi \lessgtr 1$ are depicted
in Figs. \ref{fig:N5chi0.8e1_V} and \ref{fig:N10chi1.8e1_V}.
In all the numerical calculations in this paper, 
we set the energy splitting between the two levels $\epsilon=1$.
In other words, we measure energy in units of $\epsilon$.

\subsection{Moving-frame equations with the Lipkin model}

We shall consider the basic equations (\ref{eq:mfHF})-(\ref{eq:p^2 EoCSM}) 
employing the Lipkin model.
In general cases, the basic equations may be solved by approximating 
$\partial_q \hat Q^{(1)}(q)$ with the finite difference.
With this simple model, 
the basic equations reduce to one first-order differential equation
of $Q^{(1)}(\phi)$ as shown below.
Following Ref. \cite{Holzwarth1973}, we assume that any collective
operator is written in terms of the quasispin operators $\hat J_{\pm}$
and $\hat J_{0}$.
In Approach A, the collective operators are written in terms of only
A-terms, which implies that they are written in terms of $\hat J_{\pm}$.
Noting the Hermiticity and the time-reversal symmetry of the collective
operators, one finds
\begin{align}
 \hat Q^{(1)}(q) & = Q^{(1)}(q) (\hat J_+(q) + \hat J_-(q)), \label{eq:Q1 in terms of J}\\
  B \hat P(q)    & = i BP(q) (\hat J_+(q) - \hat J_-(q)),  \label{eq:BP in terms of J}\\
 \hat Q^{(2)}(q) & = iQ^{(2)}(q) (\hat J_+(q) - \hat J_-(q)). \label{eq:Q2 in terms of J}
\end{align}
Here, $Q^{(1)}(q), BP(q)$ and $Q^{(2)}(q)$ are real numbers.
While $\hat Q^{(1)}$ is time-even, $\hat P$ and $\hat Q^{(2)}$ are time-odd. 
The canonical-variable condition \cite{Matsuo2000} determines
the normalization of $(Q^{(1)}(q), P(q))$ as follows.
\begin{align}
& \langle \phi(q)|[\hat Q^{(1)},\hat P]|\phi(q)\rangle =i
\Leftrightarrow
Q^{(1)} P=\frac{1}{2N}.
\label{eq:normalization of Q_A,P}
\end{align}

As shown in Appendix A, the derivatives of $(\hat Q^{(1)}, \hat P)$ are
written as
\begin{align}
 \partial_q \hat Q^{(1)}(q) &=\partial_qQ^{(1)}(q) (\hat J_+(q)+\hat
 J_(q)) +\frac{2}{N}\hat J_0(q), \label{eq:dqQ1 in terms of J}\\
 \partial_q \hat P(q) &=i\partial_qP(q) (\hat J_+(q)-\hat J_(q)).
\end{align}
In the ASCC method, 
as only the variation of A-type, 
$a^\dagger a^\dagger |\phi(q)\rangle$, is taken,
so here we take the variation of the form  $\hat J_+ |\phi(q)\rangle$.

We shall give the expression of the inertial mass $B(q)$ first.
%
It is given by Eq. (\ref{eq:B(q)}), and using Eqs.(\ref{eq:H in terms of tilde J0}), (\ref{eq:Q1 in terms of J}), and (\ref{eq:BP in terms of J}),
we have
\begin{align}
B(q)=-N\epsilon\sin\phi(1-\chi\cos\phi)Q^{(2)}
+2N\left[\epsilon\cos\phi+\epsilon\chi(1+\sin^2\phi)\right]\left(Q^{(1)}\right)^2.
\label{eq:B(q) with Lipkin}
\end{align}
Although we choose the coordinate system with $B(q)\equiv 1$,
we keep $B(q)$ explicitly in the expressions below.
One can always choose such a coordinate system with $B=1$ by the scale
transformation of the collective coordinate $q$.
Note that $Q^{(1)}$ and $Q^{(2)}$ are rank-1 and rank-2 contravariant
tensors, respectively.
$B(q)$ is a rank-2 contravariant tensor,
 and $P$ is a rank-1 covariant tensor (vector).

Let us move on to the equations of motion.
The moving-frame HF equation (\ref{eq:mfHF}) reads
\begin{align}
 \frac{1}{2}\epsilon\sin\phi(1-\chi\cos\phi)-\partial_qV Q^{(1)}=0.
\label{eq:mfHF with Lipkin}
\end{align}
From Eq. (\ref{eq:HF condition}), we have
\begin{align}
\frac{\partial V}{\partial \phi}
= \frac{\partial V}{\partial q}\frac{\partial q}{\partial \phi}=\frac{1}{2}N\epsilon \sin\phi \left(1-\chi \cos\phi\right).
\end{align}
By comparing this with Eq. (\ref{eq:mfHF with Lipkin}), we find
\begin{align}
 \frac{\partial \phi}{\partial q}= \frac{1}{NQ^{(1)}(q)}=2P(q).
\label{eq:dphi dq}
\end{align}
The last equality follows from the canonical-variable conditions
(\ref{eq:normalization of Q_A,P}).

Let us consider the rest of the basic equations.
First, we shall see the case where $\partial_qV=0$.
Eqs. (\ref{eq:mfRPA1 with Q2}) and (\ref{eq:dq moving-frame HF eq.})
reduce to the RPA equations:
\begin{align}
\delta \langle \phi | [\hat H, \hat Q^{(1)}]-\frac{1}{i}B\hat P|\phi\rangle=0, \label{eq:RPA1 in QP rep}\\
\delta \langle \phi |  [\hat H, \frac{1}{i}\hat BP]-BC\hat Q^{(1)} |\phi\rangle=0. \label{eq:RPA2 in QP rep}
\end{align}

Using Eqs.(\ref{eq:H in terms of tilde J0}), (\ref{eq:Q1 in terms of J}), 
and (\ref{eq:BP in terms of J}),
we have
\begin{align}
&\delta \langle \phi | [\hat H, \hat Q^{(1)}]-\frac{1}{i}B\hat P|\phi\rangle=0
\Leftrightarrow
\langle \phi | \hat J_- \left\{
 [\hat H, \hat Q^{(1)}]-\frac{1}{i}B\hat P \right\}|\phi \rangle=0 \notag \\
\Leftrightarrow
&\left[\epsilon\cos\phi +\epsilon\chi(1+\sin^2\phi)\right]Q^{(1)} -BP=0.
\label{eq:RPA1}
\end{align}
\begin{align}
&\delta \langle \phi | [\hat H, \frac{1}{i}B\hat P]-\hat Q^{(1)}|\phi\rangle=0
\Leftrightarrow
\langle \phi | \hat J_- \left( [\hat H, \frac{1}{i}B\hat P]-\hat Q^{(1)}\right)|\phi\rangle=0,\notag\\
\Leftrightarrow
&\left[\epsilon\cos\phi+\epsilon\chi\sin^2\phi -\epsilon\chi \cos^2\phi \right]BP-BC Q^{(1)}=0.
\label{eq:RPA2}
\end{align}
The RPA equations  can be written as
\begin{align}
\begin{pmatrix}
\epsilon\cos\phi+\epsilon\chi\sin^2\phi +\epsilon\chi  && -1 \\
-BC && \epsilon\cos\phi +\epsilon\chi(\sin^2\phi-\cos^2\phi)
\end{pmatrix}
\begin{pmatrix}
Q^{(1)}  \\ 
BP
\end{pmatrix}
=
\begin{pmatrix}
0 \\ 0
\end{pmatrix},
\end{align}
Here, $BC$ plays a role of the eigenfrequency squared $\omega^2$ and is given by
\begin{align}
BC=\omega^2=\left[\epsilon\cos\phi+\epsilon\chi\sin^2\phi +
 \epsilon\chi\right]
\left[\epsilon\cos\phi +\epsilon\chi(\sin^2\phi-\cos^2\phi)\right].
\end{align}
The RPA solution reads
\begin{align}
 \begin{pmatrix}
Q^{(1)}  \\ 
BP
\end{pmatrix}
=
\sqrt{\frac{B}{2N}}
\begin{pmatrix}
\left[\epsilon\cos\phi  +\epsilon\chi(1+\sin^2\phi) \right]^{-\frac{1}{2}} 
\\ 
\left[\epsilon\cos\phi  +\epsilon\chi(1+\sin^2\phi)\right]^{\frac{1}{2}} 
\end{pmatrix}.
\label{eq:Q1, P in RPA}
\end{align}
We have used the canonical-variable condition (\ref{eq:normalization of
Q_A,P}) for the normalization.
Here $\phi$ in the above expression is a solution to the HF equation  where $\partial_qV=0$.
One easily sees that $\phi=0$ and $\phi=\pi$ are always solutions to the HF equation (\ref{eq:HF condition}),
and that for $\chi>1$ there is another solution
$\phi_0:=\cos^{-1}(1/\chi)$.
For $\chi>1$,
the RPA solution at $\phi=\phi_0$ is given by
\begin{align}
 \begin{pmatrix}
Q^{(1)}  \\ 
BP
\end{pmatrix}
=
\sqrt{\frac{B}{2N}}
\begin{pmatrix}
\left[2\epsilon\chi\right]^{-\frac{1}{2}} 
\\ 
\left[2\epsilon\chi\right]^{\frac{1}{2}} 
\end{pmatrix}.
\end{align}
The solutions at $\phi=0$ and $\phi=\pi$ are
\begin{align}
 \begin{pmatrix}
Q^{(1)}  \\ 
BP
\end{pmatrix}
=
\sqrt{\frac{B}{2N}}
\begin{pmatrix}
\left[\epsilon(\chi\pm 1)\right]^{-\frac{1}{2}} 
\\ 
\left[\epsilon(\chi\pm 1~\right]^{\frac{1}{2}} 
\end{pmatrix},
\label{eq: RPA solution at 0, pi}
\end{align}
where $+$ and $-$ correspond to $\phi=0$ and $\phi=\pi$, respectively.

Next, noting that 
$[\hat Q^{(1)},\hat Q^{(2)}]$ is a B-term,
one finds that
Eq. (\ref{eq:p^2 EoCSM}) is independent
of $\partial_qV$.
From Eq. (\ref{eq:p^2 EoCSM}), we obtain
\begin{align}
 \epsilon \chi \cos\phi \sin \phi \left(Q^{(1)}\right)^2
+B D_q Q^{(1)} 
-\frac{1}{2}\left[\epsilon\cos\phi +\epsilon\chi(\sin^2\phi-\cos^2\phi)\right]Q^{(2)}=0,
\label{eq:2nd-order EoCSM with Lipkin}
\end{align}
where $D_q Q^{(1)}=\partial_qQ^{(1)}+\Gamma (q)Q^{(1)}$.
This equation gives the relation between $\partial_q Q^{(1)}$ and $Q^{(2)}$.
(At this point, the value of 
$\partial_q Q^{(1)}$ at the HF equilibrium point is unknown, so is $Q^{(2)}$.)
The partial derivative
$\partial_qQ^{(1)}$ can be rewritten as
\begin{align}
& \partial_q Q^{(1)} =\frac{\partial Q^{(1)}}{\partial
 \phi}\frac{\partial \phi}{\partial q}
=2P\frac{\partial Q^{(1)}}{\partial\phi}
=\frac{1}{NQ^{(1)}}\frac{\partial Q^{(1)}}{\partial\phi},
\label{eq:dqQ1}
\end{align}
using Eq. (\ref{eq:dphi dq}),
so if we choose the coordinate system with
$B(q)\equiv 1$ and can express 
$Q^{(2)}$ in terms of  $Q^{(1)}$,
then Eq. (\ref{eq:2nd-order EoCSM with Lipkin}) reduces 
to a differential equation of $Q^{(1)}$ with respect to $\phi$.

We shall move on to the case where $\partial_qV \neq 0$. 
Then, as long as Eq. (\ref{eq:p^2 EoCSM}) is adopted as one of the basic equations,
the set of Eqs. (\ref{eq:mfRPA1 with Q2}) and (\ref{eq:dq moving-frame HF eq.})
are equivalent to the moving-frame RPA equations  and
leads to 
\begin{align}
& \left[\epsilon\cos\phi +\epsilon\chi(1+\sin^2\phi)\right]
 Q^{(1)}-BP -\partial_qV Q^{(2)}=0,
\label{eq:moving-frame RPA1 in App. A w Lipkin} 
\end{align}
\begin{align}
& \left[\epsilon\cos\phi +\epsilon\chi(\sin^2\phi-\cos^2\phi)\right]BP 
-BCQ^{(1)}\notag\\
&-\frac{1}{2} \left[\epsilon\cos\phi +\epsilon\chi(\sin^2\phi-\cos^2\phi)\right]
\partial_qV Q^{(2)}
+\epsilon\chi\cos\phi\sin\phi\partial_q V\left( Q^{(1)}\right)^2
=0.
\label{eq:moving-frame RPA2 in App. A w Lipkin}
\end{align}

We shall express $BC=\omega^2$ as a function of $Q^{(1)}$ or
equivalently $P$ [See Eq. (\ref{eq:normalization of Q_A,P})].
By eliminating $\partial_qV\hat Q^{(2)} $ from
Eq. (\ref{eq:moving-frame RPA2 in App. A w Lipkin})
with use of Eq. (\ref{eq:moving-frame RPA1 in App. A w Lipkin}),
we find
\begin{align} 
&
BC=3N\left[\epsilon\cos\phi +\epsilon\chi(\sin^2\phi-\cos^2\phi)\right]BP^2 
\notag\\
&-\frac{1}{2}
\left[\epsilon\cos\phi +\epsilon\chi(\sin^2\phi-\cos^2\phi)\right] 
\left[\epsilon\cos\phi +\epsilon\chi(1+\sin^2\phi)\right]
\notag\\ 
&
+\frac{1}{2}\epsilon^2\chi\cos\phi\sin^2\phi(1-\chi\cos\phi)
\label{eq:BCQ1 in App. A}
\end{align}
Here we have used the moving-frame HF equation (\ref{eq:mfHF with Lipkin}) and the
canonical-variable condition (\ref{eq:normalization of Q_A,P}).
Now $BC$ is written in terms of $P$ and is readily rewritten in terms of $Q^{(1)}$.
Eq. (\ref{eq:BCQ1 in App. A}) can be also obtained by calculating
the product of $B$ and $C$ through Eqs. (\ref{eq:M(phi) in terms of Q1
and P}), (\ref{eq:dqQ1 from derivative of mfHF}), 
and (\ref{eq:definition of C}) in the $\phi$ space.

Let us remove $Q^{(2)}$
from Eq. (\ref{eq:2nd-order EoCSM with Lipkin}).
With the moving-frame RPA equation (\ref{eq:moving-frame RPA1 in App. A
w Lipkin})$\times Q^{(1)}$ and the moving-frame HF equation (\ref{eq:mfHF with Lipkin}), we obtain
\begin{align}
 Q^{(2)}=
\frac{ 2N\left[\epsilon\cos\phi+\epsilon\chi(1+\sin^2\phi)\right]\left(Q^{(1)}\right)^2-B}
{N\epsilon\sin\phi(1-\chi\cos\phi)}.
\label{eq:Q2 from Q1}
\end{align}
Note that this is the indeterminate form of $0/0$ when $\partial_qV=0$.
Eq .(\ref{eq:Q2 from Q1}) can be also
obtained 
from the expression of the
inertial mass (\ref{eq:B(q) with Lipkin}).

By substituting Eq.
(\ref{eq:Q2 from Q1})
into Eq. (\ref{eq:2nd-order EoCSM with Lipkin}),
and setting $B\equiv 1$,
we obtain the differential equation of
$Q^{(1)}$ below.
\begin{align}
&\partial_{\phi} Q^{(1)}\notag\\
=&
\frac{1}{2}
\left[\cos\phi+\chi(\sin^2\phi-\cos^2\phi)\right]
\left\{
2N
\left[\epsilon\cos\phi+\epsilon\chi(1+\sin^2\phi)\right]
\left(Q^{(1)}\right)^2
-1
\right\}
\frac{Q^{(1)}}{\sin\phi(1-\chi\cos\phi)}\notag\\
&
-N\epsilon\chi\cos\phi\sin\phi \left(Q^{(1)}\right)^3.
\label{eq:dqQ1 from derivative of mfHF}
\end{align}
Now all one has to do is  to integrate the above differential 
equation as an initial-value problem.
We choose $\phi=0$ as the initial point.
From the symmetry of the system,
we can assume that $Q^{(1)}(\phi)$ is symmetric under the reflection about $\phi=0$
and differentiable at $\phi=0$, and then $\partial_\phi Q^{(1)}(\phi=0)=0$.
From the $O(p^2)$ equation of CS (\ref{eq:2nd-order EoCSM with Lipkin}) 
and $\partial_\phi Q^{(1)}(\phi=0)=0$,
it follows that $Q^{(2)}(\phi)$ vanishes at $\phi=0$.

The relation between $q$ and $\phi$ can be also obtained
from Eq. (\ref{eq:dphi dq}) and 
we find
\begin{align}
 q=N\int_{\phi_0}^{\phi} Q^{(1)}(\phi^\prime) d\phi^\prime.
\label{eq:q(phi)}
\end{align}
We have chosen the origin $q=0$ such that
$\phi=\phi_0=\cos^{-1}(1/\chi)$ at $q=0$.

Before ending this subsection, we shall give a remark.
We have seen that the $q$-derivative of the $O(1)$ equation of CS
can be used as one of the basic equations independent of the moving-frame
HF \& RPA equations.
One may wonder if also the $q$-derivative of the $O(p)$ equation of CS
may be used as a basic equation.
However, there appears no independent equation,
if one differentiates the $O(p)$ equation of CS with respect to $q$.
In the case of the Lipkin model,
the $q-$derivative of the $O(p)$ equation of CS (\ref{eq:mfRPA1 with Q2}) leads to
\begin{align}
& 2\epsilon\chi\cos\phi\sin\phi Q^{(1)}P 
+\left[\epsilon\cos\phi+\epsilon\chi(1+\sin^2\phi)\right]\partial_qQ^{(1)}
-\frac{1}{N}\epsilon\sin\phi(1-\chi\cos\phi) \notag\\
-&\left(B\partial_qP+\partial_qV \partial_q Q^{(2)} \right)
-C Q^{(2)}
+\partial_qB 
\left(\frac{1}{2B}
\partial_qV Q^{(2)}
-\partial_qB P
\right)
=0,
\label{eq: dq moving-frame RPA of O(p) in App. A w Lipkin}
\end{align}
which can be simplified with $B\equiv 1$ and the canonical-variable
condition (\ref{eq:normalization of Q_A,P}) as below.
\begin{align}
& 
\left[\epsilon\cos\phi+\epsilon\chi(1+\sin^2\phi)
+\frac{1}{2N(Q^{(1)})^2}
\right]\partial_qQ^{(1)}\notag\\
+&\frac{1}{N}\epsilon\sin\phi(2\chi\cos\phi-1)
-CQ^{(2)}-\partial_qV \partial_q Q^{(2)} 
=0.
\label{eq:dq 1st-order EoCSM with Lipkin}
\end{align}
%
The above equation (\ref{eq:dq 1st-order EoCSM with Lipkin}) is also directly obtained
by differentiating Eq. (\ref{eq:moving-frame RPA1 in App. A w Lipkin})
with respect to $q$ and using Eqs. (\ref{eq:dphi dq}) and (\ref{eq:normalization of Q_A,P}),
so it is not independent of those equations.
Thus, Eq. (\ref{eq:dq 1st-order EoCSM with Lipkin}) gives no new
condition to the set of the $O(1)$ and $O(p)$ equations of CS. 
Which equations are independent of one another depends on the form of the state vector.



\subsection{Collective mass and Schr\"odinger equation}
When we choose the coordinate system such that $B(q)=1$, the collective Hamiltonian
reads
\begin{align}
 \mathcal{H}=\frac{1}{2}\dot q^2 + V(q).
\end{align}
We shall rewrite the kinetic energy in terms of $\phi$.
\begin{align}
 T&=\frac{1}{2}\dot q^2 =\frac{1}{2}\left(\frac{\partial q}{\partial
 \phi} \right)^2 \dot \phi^2  :=\frac{1}{2}M(\phi)\dot \phi^2.
\end{align}
Here, the inertial mass $M(\phi)$ can be written as
\begin{align}
 M(\phi)= \left(\frac{\partial q}{\partial \phi} \right)^2 = N^2 (Q^{(1)})^2=\frac{1}{4P^2},
\label{eq:M(phi) in terms of Q1 and P}
\end{align}
where we have used Eq. (\ref{eq:dphi dq}).



We solve the collective Schr\"odinger equation in the $\phi$ space rather than the $q$ space
because it is more convenient for the comparison between the calculations
with and without $\hat Q^{(2)}$.
The classical collective Hamiltonian is written in terms of $\phi$ as
\begin{align}
 \mathcal{H}=\frac{1}{2}M(\phi)\dot \phi^2 +V(\phi).
\end{align}
The quantized Hamiltonian is given by
\begin{align}
 \hat H_{\rm coll} =\hat T +V = -\frac{1}{2}\Delta +V(\phi),
\end{align}
with 
\begin{align}
 \Delta&=\frac{1}{\sqrt{M}}\partial_\phi \sqrt{M}\frac{1}{M}\partial_\phi\notag\\
&=\frac{1}{M}\partial_\phi^2-\frac{1}{2}\frac{1}{M^2} \left(\partial_\phi M \right)\partial_\phi
=B\partial_\phi^2+\frac{1}{2}\partial_\phi B \partial_\phi,
\end{align}
with $B(\phi)=M^{-1}(\phi)$.
For this Hamiltonian, we solve the collective Schr\"odinger equation under
the boundary condition explained in the next subsection.


\subsection{Solution in the case without $\hat Q^{(2)}$}
\label{sec: soltion without Q2}
In the ASCC theory without $\hat Q^{(2)}$,
the equations of motion are given by the moving-frame HF equation
(\ref{eq:mfHF with Lipkin}) and the moving-frame RPA equations with
$Q^{(2)}$ omitted,
\begin{align}
& \left[\epsilon\cos\phi +\epsilon\chi(1+\sin^2\phi)\right]
 Q^{(1)}-BP =0,
\label{eq:moving-frame RPA1 without Q2 w Lipkin} 
\end{align}
\begin{align}
& \left[\epsilon\cos\phi +\epsilon\chi(\sin^2\phi-\cos^2\phi)\right]BP 
-BCQ^{(1)}
+\epsilon\chi\cos\phi\sin\phi\partial_q V\left( Q^{(1)}\right)^2
=0.
\label{eq:moving-frame RPA2 without Q2 w Lipkin}
\end{align}
From Eq. (\ref{eq:moving-frame RPA1 without Q2 w Lipkin}) and the
canonical-variable condition (\ref{eq:normalization of Q_A,P}),
we obtain
\begin{align}
 \begin{pmatrix}
Q^{(1)}  \\ 
BP
\end{pmatrix}
=
\sqrt{\frac{B}{2N}}
\begin{pmatrix}
\left[\epsilon\cos\phi  +\epsilon\chi(1+\sin^2\phi) \right]^{-\frac{1}{2}} 
\\ 
\left[\epsilon\cos\phi  +\epsilon\chi(1+\sin^2\phi)\right]^{\frac{1}{2}} 
\end{pmatrix}.
\label{eq:Q1, P in ASCC without Q2}
\end{align}
We shall call the moving-frame RPA equations without the curvature term
[the last term in Eq. (\ref{eq:moving-frame RPA2 without Q2 w Lipkin})]
the local RPA (LRPA) equations.
As $(Q^{(1)},BP)$ are determined from Eq. 
(\ref{eq:moving-frame RPA1 without Q2 w Lipkin}) and the
canonical-variable condition (\ref{eq:normalization of Q_A,P}),
the solution $(Q^{(1)},BP)$ of the moving-frame RPA equations 
coincides with that to the LRPA solution.
Then, the inverse collective mass is given by
\begin{align}
B(\phi)= M^{-1}(\phi)
=\frac{2}{N}\epsilon \left[\cos\phi+\chi\left(1+\sin^2\phi\right)\right],
\label{eq:B(phi) without Q2}
\end{align}
which coincides with the ATDHF mass (41) and the GCM mass (72) in Ref.~\cite{Holzwarth1973}.
When $\chi<1$, $B$ is positive for sufficiently small $\phi \in [0,\pi]$, 
but $B$ becomes 0 at a certain point, beyond which 
$(\hat Q^{(1)},\hat P)$ is no longer Hermitian and $B< 0$.
We shall denote the point at which $B=0$ by $\phi_{\rm max}\in (0,\pi)$.
We solve the collective Schr\"odinger equation with the boundary
condition for the collective wave functions 
to vanish outside the region $(-\phi_{\rm max} ,\phi_{\rm max})$,
because the potential energy is sufficiently high there.
When $\chi>1$, $B$ is always positive.
Then, we solve the collective Schr\"odinger equation
with the periodic boundary condition.
We employ these boundary conditions similarly in the case with
$Q^{(2)}$ included.


The eigenfrequency $\omega^2=BC$ of the moving-frame RPA equations without $\hat Q^{(2)}$
is obtained from Eq. (\ref{eq:moving-frame RPA2 without Q2 w Lipkin}) as
\begin{align}
 BC&=
\left[\epsilon\cos\phi +\epsilon\chi(1+\sin^2\phi)\right]
\left[\epsilon\cos\phi +\epsilon\chi(\sin^2\phi-\cos^2\phi)\right]\notag\\
&+\frac{1}{2}\epsilon^2\chi\sin^2\phi \cos\phi (1-\chi\cos\phi),
\label{eq:BC in ASCC without Q2}
\end{align}
and $\partial_q V$ is also obtained analytically
\begin{align}
 \partial_qV=
\sqrt{\frac{N}{2B}}\epsilon\sin\phi(1-\chi\cos\phi)
\left[\epsilon\cos\phi  +\epsilon\chi(1+\sin^2\phi)\right]^{\frac{1}{2}}.
\end{align}

\section{Numerical results}

We compare the numerical results obtained by solving
Eq. (\ref{eq:dqQ1 from derivative of mfHF}) with those 
obtained from the conventional ASCC equations without $\hat Q^{(2)}$.
For the symmetry of the system, we solve Eq. (\ref{eq:dqQ1 from
derivative of mfHF}) in the region $[0,180^\circ]$ as an initial-value problem,
starting from $\phi=0$.

\begin{figure}[tbp]
\centering
\includegraphics[width=0.95\textwidth]{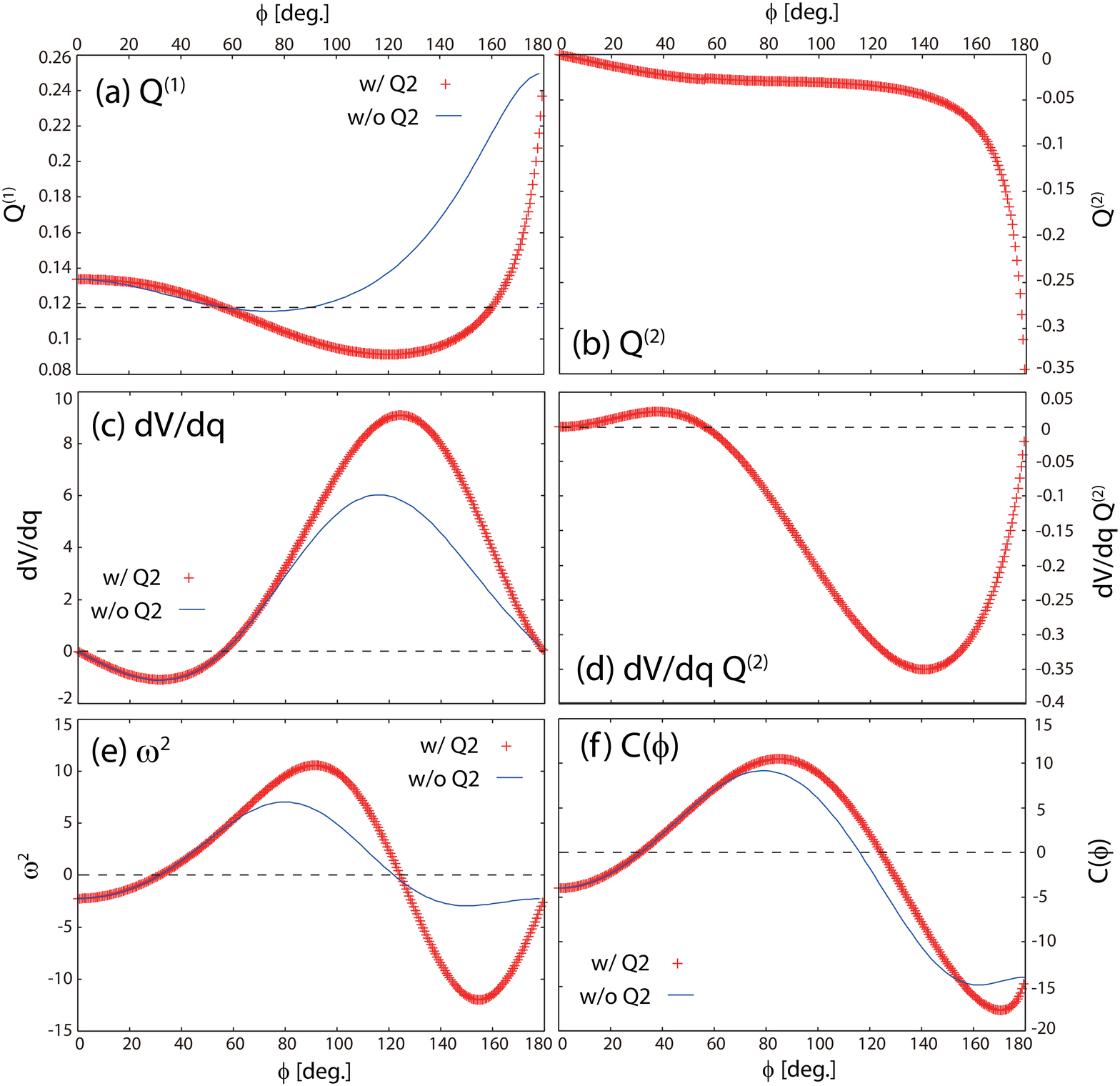}
 \caption{
Numerical results obtained by solving Eq. (\ref{eq:dqQ1 from derivative of mfHF})
for $N=10, \chi=1.8, \epsilon=1.0$ plotted as function of $\phi$ 
:
(a) $Q^{(1)}$, (b) $Q^{(2)}$, (c) $\partial_qV$, (d)
 $\partial_qVQ^{(2)}$,
(e) $\omega^2$, and (f) $C(\phi)$.
For (a) $Q^{(1)}$, (c) $\partial_qV$, (e) $\omega^2$ and (f) $C(\phi)$, the results obtained by
 solving the ASCC equations without $Q^{(2)}$ are shown 
with the solid line for comparison. The dashed line in the panel (a)
indicates the value of the RPA solution (\ref{eq:Q1, P in RPA}) at $\phi=\phi_0$.
(See also Fig. \ref{fig:N10chi1.8e1_phi_vs_Q1_AppA_from0_rk300_mag}.)
}
\label{N10chi1.8e1_AppA_from0_rk300}
 \begin{center}
\includegraphics[width=0.5\textwidth]{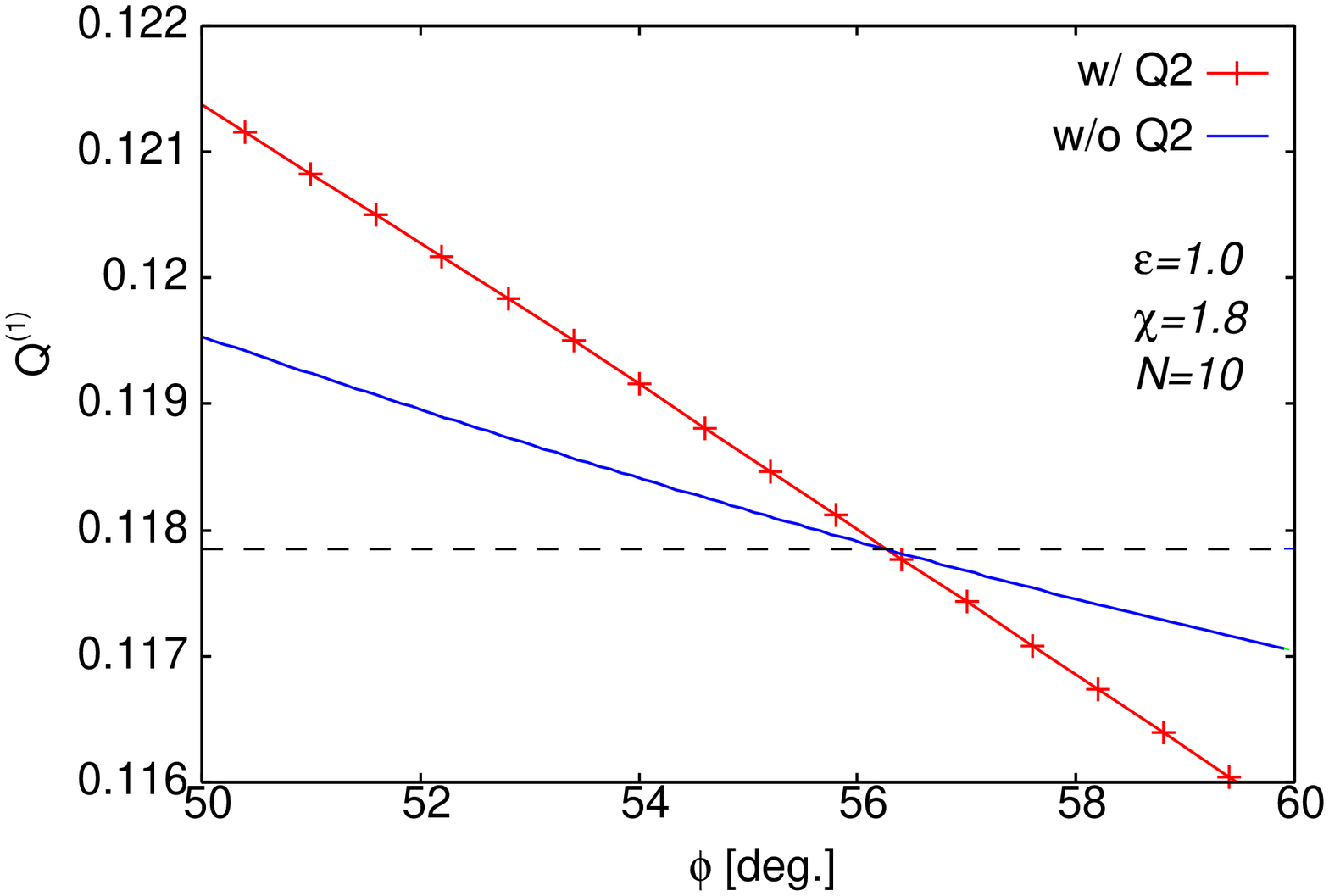}
 \end{center}
 \caption{Same as Fig. \ref{N10chi1.8e1_AppA_from0_rk300}(a), but
magnified around $\phi=\phi_0$.
The dashed line 
indicates the value of the RPA solution (\ref{eq:Q1, P in RPA}) at $\phi=\phi_0$.
At $\phi=\phi_0$, $\hat Q^{(1)}$ coincides the RPA solution.}
\label{fig:N10chi1.8e1_phi_vs_Q1_AppA_from0_rk300_mag}
\end{figure}

As an example of the calculations for $\chi>1$, we show the calculated results
for $\chi=1.8, N=10, \epsilon=1$ in Fig. \ref{N10chi1.8e1_AppA_from0_rk300}.
We depict (a) $Q^{(1)}$, (b) $Q^{(2)}$, (c) $\partial_qV$, (d) $
\partial_qVQ^{(2)}$, 
(e) $\omega^2$
and (f) $C$ as functions of $\phi$.
When $Q^{(2)}$ is not included,
$Q^{(1)}$, shown in Fig. \ref{N10chi1.8e1_AppA_from0_rk300}(a), 
is obtained analytically as seen in the previous section. 
It coincides with the RPA solutions at the potential extrema $\phi=0, \phi_0$ and $\pi$.
One can see that, also with $\hat Q^{(2)}$, the calculated result 
coincides with the RPA solutions at the potential extrema. 
(In Fig. \ref{fig:N10chi1.8e1_phi_vs_Q1_AppA_from0_rk300_mag}, 
the magnified figure around $\phi=\phi_0=56.3^\circ$ is shown).
It is also noteworthy that, for $\phi \lesssim \phi_0$, both of the
calculations give similar results, but for $\phi \gtrsim \phi_0$, the
deviation between the two becomes larger.
[We shall give a minor remark. 
At $\phi=180^\circ$, the differentiability of $Q^{(1)}$ seems to be
broken if we assume $B(\phi)$ is periodic. 
However, this is not a serious problem because the potential
energy is sufficiently high and the collective wave function almost
vanishes there.
We also integrated Eq. (\ref{eq:dqQ1 from derivative of mfHF}) 
from $\phi=180^\circ$ with the boundary
condition $\partial_\phi Q^{(1)}=0$ as in the case from $\phi=0$.
The obtained solution coincides with the RPA solutions at the
potential extrema, but in turn the differentiability at $\phi=0$ is broken.
Therefore, we adopt here the solution obtained by the integration from
$\phi=0$.
]

In Fig. \ref{N10chi1.8e1_AppA_from0_rk300}(b), $Q^{(2)}$ is plotted. 
It vanishes at $\phi=0$ and rapidly decreases as $\phi$ approaches to
$180^\circ$. However, it does not diverge there but converges to a
finite value.
$Q^{(2)}$ is involved in the equations of motion in the form of
$\partial_qVQ^{(2)}$, so we plot $\partial_qVQ^{(2)}$  in Fig. \ref{N10chi1.8e1_AppA_from0_rk300}(d). As $\partial_q V=0$ at the potential
extrema $\phi=0^\circ, \phi_0$,  and $180^\circ$ and $Q^{(2)}$ is always finite, 
their product vanishes at $\phi=0, \phi_0$, and $180^\circ$. Thus,
the equations of motion reduce to the HF \& RPA equations there.
[Note that, $\phi=0, 180^{\circ}$ are (local) potential maxima, although
the RPA equations are usually solved at the potential minimum.] 

Figs. \ref{N10chi1.8e1_AppA_from0_rk300}(e) and \ref{N10chi1.8e1_AppA_from0_rk300}(f) display
the eigenfrequency squared $\omega^2$ and the potential curvature $C$ as
functions of $\phi$, respectively.
In the calculation with $\hat Q^{(2)}$ included, $\omega^2$, which is
calculated through Eq. (\ref{eq:BCQ1 in App. A}), 
coincides with the product of the curvature $C$ and the inverse inertial mass $B$,
i. e., $\omega^2=BC$.
However, it is not the case when $\hat Q^{(2)}$ is ignored. 
(It is seen that the position of the zero of $\omega^2$ around
$120^\circ$ is different from that of $C$.)
This will be investigated in the next section.

\begin{figure}[tbph]
 \begin{center}
\includegraphics[width=0.5\textwidth]{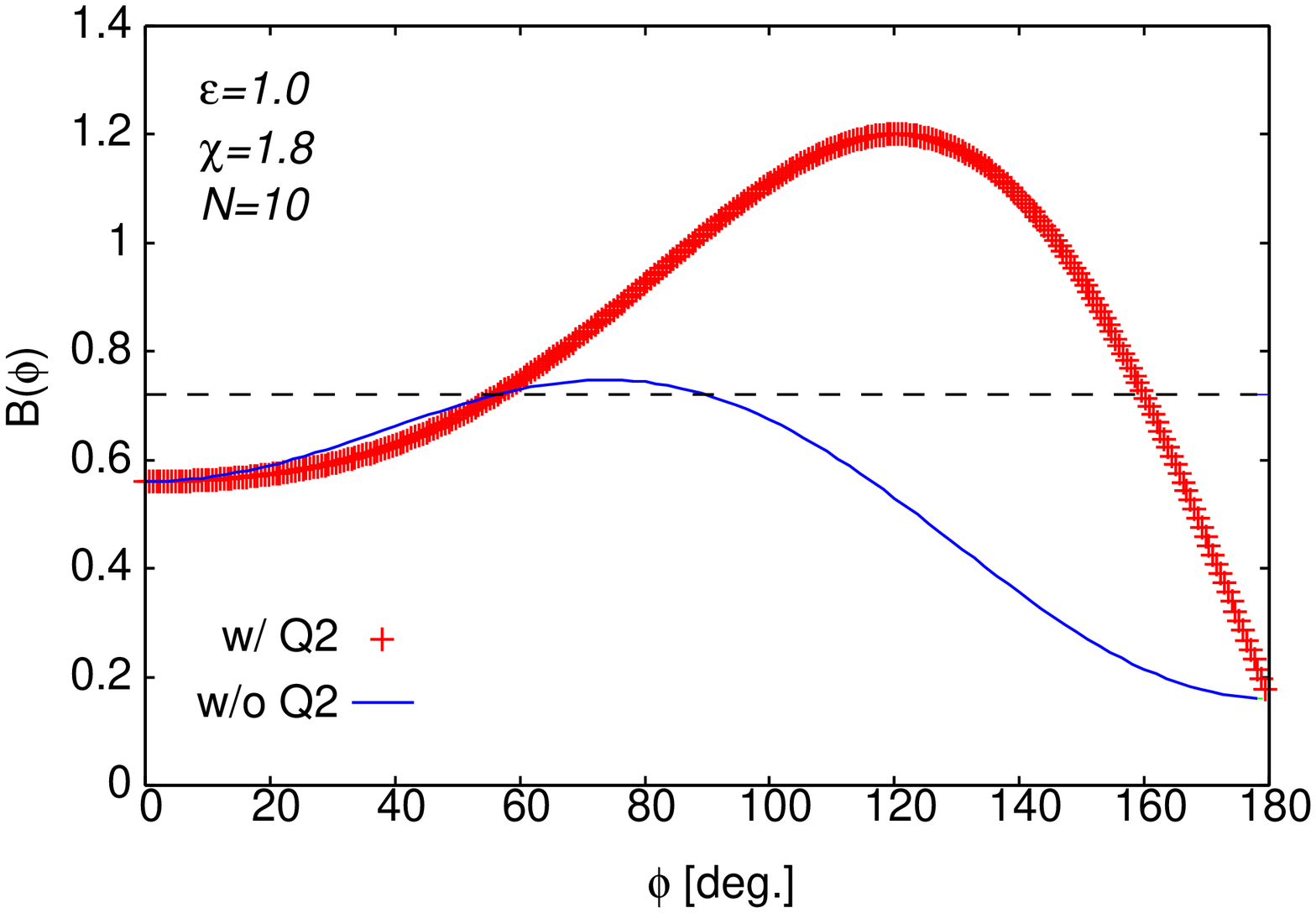}
 \end{center}
 \caption{Inverse inertial mass
$B(\phi)$ obtained by solving
(\ref{eq:dqQ1 from derivative of mfHF}) for $N=10, \chi=1.8,
 \epsilon=1.0$.  
The inverse inertial mass obtained without $\hat Q^{(2)}$
is also shown for comparison.
}
\label{fig:B(phi) for N=10}
%
 \begin{center}
\includegraphics[width=0.5\textwidth]{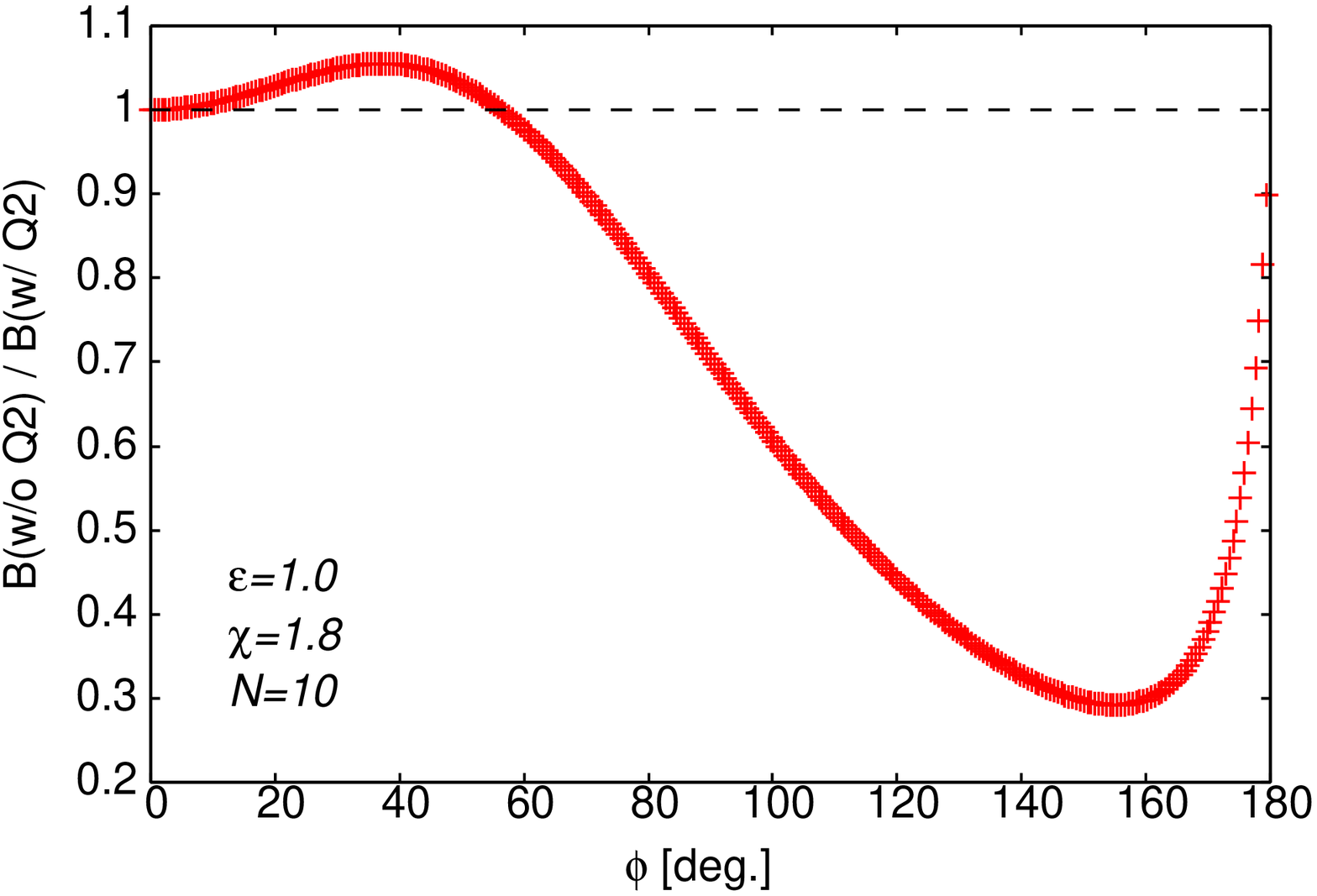}
 \end{center}
 \caption{
Ratio of the 
inverse inertial mass
$B(\phi)$ calculated without $Q^{(2)}$ to that with $Q^{(2)}$
for $N=10, \chi=1.8,  \epsilon=1.0$.
}
\label{fig:ratio of B(wo Q2) to B(w Q2) for N=10} 
\end{figure}

The inverse inertial mass $B(\phi)$ calculated with $Q^{(2)}$
is shown in Fig. \ref{fig:B(phi) for N=10},
in comparison with that calculated without $Q^{(2)}$.   
In Fig. \ref{fig:ratio of B(wo Q2) to B(w Q2) for N=10},
the ratio of the inverse inertial mass calculated without $\hat Q^{(2)}$ to that
with $\hat Q^{(2)}$, 
\begin{align}
  \frac{ B(\phi)\,\,{\rm w/o}\,\,\hat Q^{(2)}}{ B(\phi)\,\,{\rm w/}\,\,\hat Q^{(2)}}
= \frac{ M(\phi)\,\,{\rm w/}\,\,\hat Q^{(2)}}{ M(\phi)\,\,{\rm w/o}\,\,\hat Q^{(2)}},
\end{align}
is plotted.
Reflecting the difference in $Q^{(1)}$, 
while for $\phi \lesssim \phi_0$, the difference between the two is not so large,
it becomes more significant for $\phi \gtrsim \phi_0$.
There, the inverse inertial mass with $\hat Q^{(2)}$ is larger than that without
$\hat Q^{(2)}$, and
depending on $\phi$ or equivalently on the collective coordinate $q$, 
the difference can be by a factor of 3 approximately.
The difference of the inertial masses becomes more important
as the excitation energy increases, because the component of the
collective wave function increases in the region with $\phi \gtrsim \phi_0$, 
where the potential energy is high and the mass difference becomes
larger.

Table \ref{table:mfRPA excitation energy RK300} displays
the comparison of the excitation energies calculated including $\hat Q^{(2)}$
with those calculated without $\hat Q^{(2)}$ and the exact
solution for $N=10, \chi=1.8$.
For a first few excited states, the difference between 
the two calculated results is small (by several percents), and
both of the calculations are in  good agreement with the 
exact solution, although the excitation energy of the first excited
state is somewhat overestimated.
With increasing the excitation energy, the difference between the two
calculations becomes larger and amounts to $10-20$\%.
Without $\hat Q^{(2)}$, 
the deviation from the exact solution become larger with increasing the
excitation energy, while, with $\hat Q^{(2)}$, the deviation stays relatively
small. 
The fourth and fifth columns in Table \ref{table:mfRPA excitation energy RK300}
show the deviations of the calculated excitation energies from the exact
solution.
Although the calculation without $\hat Q^{(2)}$ gives a better agreement
with the exact solution for the first excited state, 
for all the other states, the calculation with $\hat Q^{(2)}$ included 
gives a better agreement.

\begin{figure}[tbp]
 \begin{center}
\includegraphics[width=0.6\textwidth]{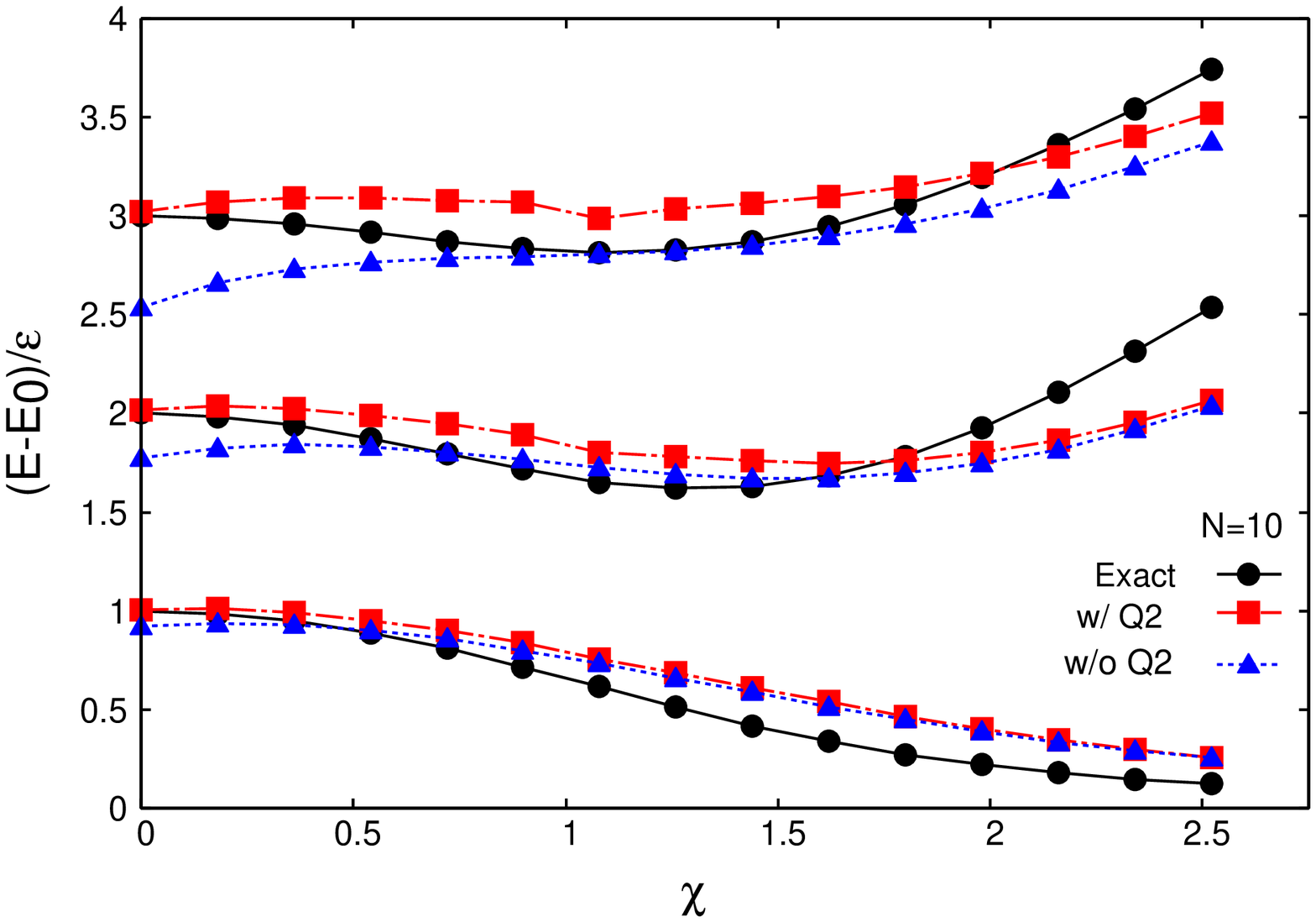}
 \end{center}
 \caption{Excitation energies of the first, second, and third excited
states for $N=10$ as functions of $\chi$.
The calculations with and without $\hat Q^{(2)}$ are compared with the
 exact solution.
}
\label{fig:Ex1-3_N10e1_dqmfHF}
%
 \begin{center}
\includegraphics[width=0.6\textwidth]{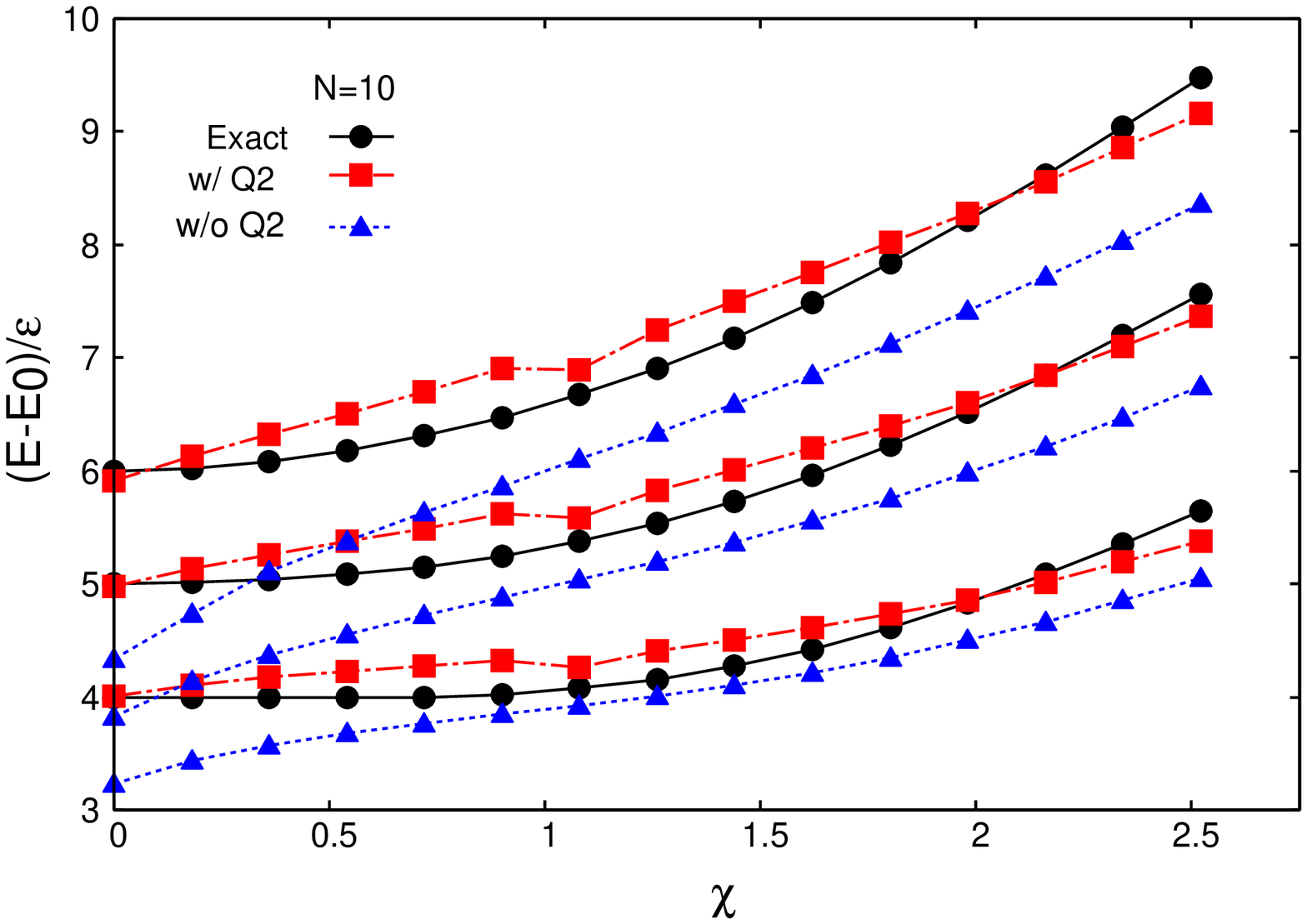}
 \end{center}
 \caption{
Same as Fig. \ref{fig:Ex1-3_N10e1_dqmfHF} but for the fourth, fifth, and
sixth excited states.
}
\label{fig:Ex4-6_N10e1_dqmfHF}
\end{figure}

Fig. \ref{fig:Ex1-3_N10e1_dqmfHF}
shows the excitation energies of the first three excited states 
as functions of $\chi$ with $N=10$.
Although the calculation with $\hat Q^{(2)}$ does not always
give a better agreement with the exact solution than that without $\hat Q^{(2)}$,
as a whole, 
the agreement with the exact solution is rather good for both of the
calculations with and without $\hat Q^{(2)}$.

Fig. \ref{fig:Ex4-6_N10e1_dqmfHF} displays the results for the next three,
the fourth, fifth and sixth excited states.  
Without $\hat Q^{(2)}$, as excitation energy increases, 
the calculated excitation energies start to deviate from the exact solution,
and
the excitation energies are systematically underestimated.
On the other hand,
the excitation energies calculated with $\hat Q^{(2)}$ are still in good
agreement with the exact solution.
These results suggest that the role of the inertial mass becomes more important 
as the excitation energy increases.

\begin{figure}[tbp]
 \begin{center}
\includegraphics[width=0.6\textwidth]{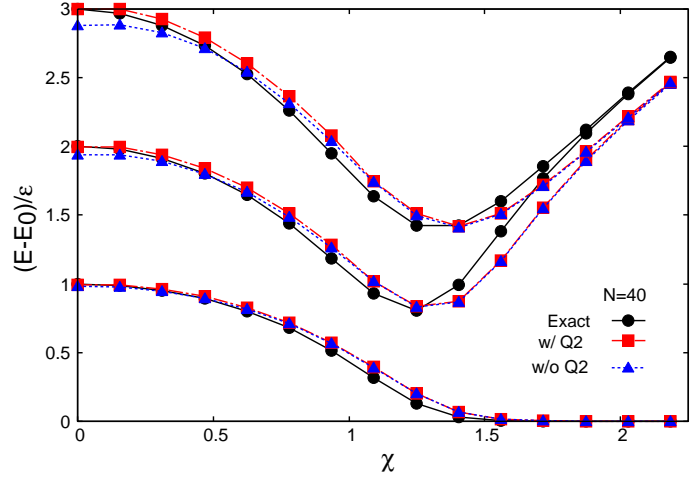}
 \end{center}
 \caption{
Same as Fig. \ref{fig:Ex1-3_N10e1_dqmfHF} but for $N=40.$
}
\label{fig:Ex1-3_N40e1_dqmfHF}
\end{figure}

\begin{figure}[tbp]
 \begin{center}
\includegraphics[width=0.6\textwidth]{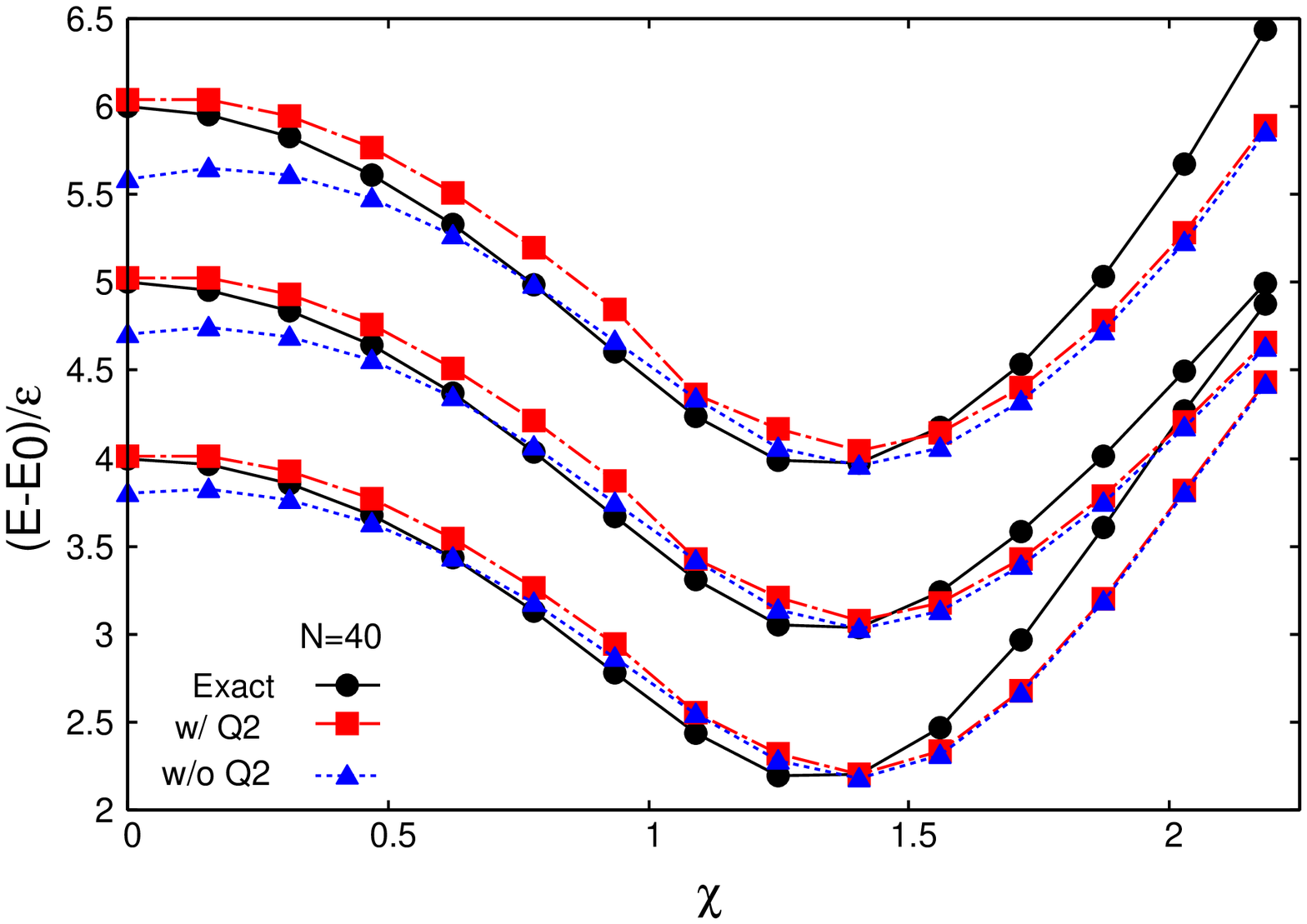}
 \end{center}
 \caption{
Same as Fig. \ref{fig:Ex1-3_N40e1_dqmfHF} but for the fourth, fifth, and
sixth excited states.
}
\label{fig:Ex4-6_N40e1_dqmfHF}
\end{figure}

We performed the calculation with a larger particle number $N=40$
similarly, and
the results are  shown in Figs. \ref{fig:Ex1-3_N40e1_dqmfHF} and \ref{fig:Ex4-6_N40e1_dqmfHF}.
As seen in the previous section, without $\hat Q^{(2)}$,
$B(\phi)$ is obtained analytically and is proportional to $N^{-1}$,
whereas the collective potential energy is proportional to $N$.
If the $N$ dependence of the inertial mass in the case with $\hat Q^{(2)}$ 
is similar to that without $\hat Q^{(2)}$, 
with increasing $N$, the potential energy becomes larger relative to the
kinetic energy,
and the difference of the inertial mass between
the two calculations becomes less important.
Actually, for $N=40$, the two calculations give similar results
and both are in good agreement with the exact solution.

\begin{table}[tbp]
\begin{center}
\begin{tabular}{cccccccc} \hline \hline
          & exact &$E$(w/o $Q^{(2)}$) & $E$ (w/ $Q^{(2)}$)  & $\Delta E$ (w/o
 $Q^{(2)}$)  &  $\Delta E$ (w/ $Q^{(2)}$) &  $\delta E/E[\%]$ \\
\hline
 $E_{ 0}$ &            0   &  0       &    0    & 0           &  0       &          \\ 
 $E_{ 1}$ &        0.2720  &  0.4482  & 0.4664	& 0.1762      &  0.1944  & 4.1     \\  
 $E_{ 2}$ &        1.784   &  1.695   & 1.760	& -0.0893     & -0.0241  & 3.8     \\  
 $E_{ 3}$ &        3.058   &  2.959   & 3.142	& -0.0988     &  0.0843  & 6.2     \\  
 $E_{ 4}$ &        4.611   &  4.350   & 4.718	& -0.2616     &  0.1065  & 8.4     \\  
 $E_{ 5}$ &        6.224   &  5.759   & 6.382	& -0.4649     &  0.1581  & 12    \\  
 $E_{ 6}$ &        7.836   &  7.124   & 7.995	& -0.7116     &  0.1587  & 12    \\  
 $E_{ 7}$ &        9.389   &  8.399   & 9.703	& -0.9897     &  0.3140   & 16    \\  
 $E_{ 8}$ &        10.66   &  9.416   & 10.46	& -1.247      & -0.2080   & 11    \\  
 $E_{ 9}$ &        12.18   &  10.62   & 12.97	& -1.569      &  0.7970   & 22    \\  
 $E_{ 10}$&        12.45   &  10.80   & 12.99   & -1.644      &  0.5420   & 20    \\  
\hline	           
\end{tabular}
\end{center}
\caption{
Excitation energies calculated with $Q^{(2)}$ for $\epsilon=1, \chi=1.8, N=10$
in comparison with those calculated without $Q^{(2)}$ and the exact solution.
The first, second, and third columns show the excitation energies of 
the exact solution, the calculation without $\hat Q^{(2)}$, and with
 $\hat Q^{(2)}$, respectively.
The fourth and fifth columns show the deviation from the exact solution
for the excitation energies calculated without and with $Q^{(2)}$, respectively.
All the energies are in units of $\epsilon (=1)$.
The rightmost column shows the percentage of the excitation-energy
increase  by $\hat Q^{(2)}$ :
$[E({\rm w/}\,\, Q^{(2)})-E({\rm w/o}\,\, Q^{(2)})]/E({\rm w/o}\,\, Q^{(2)})\times 100$.
}
\label{table:mfRPA excitation energy RK300}
\end{table}

\section{Basic equations in the case without higher-order operators}

In the previous section, 
we determined $\hat Q^{(2)}$ adopting  the $q$-derivative of the $O(1)$
equation of CS as well as the moving-frame HF \& RPA equations.
In other words, we employed 
the $O(1), O(p)$ and $O(p^2)$ equations of CS 
and the  $q$-derivative of the $O(1)$ equation of CS as independent basic equations.
It was shown that, with these equations, one can determine $\hat
Q^{(2)}$, with which the agreement with the exact solution was improved
in the Lipkin model.

One question may arise here. 
In the conventional ASCC method with only the first-order operator
$Q^{(1)}$ included, the moving-frame RPA equation of $O(p^2)$ is adopted 
as one of the basic equations, which is derived from the $O(p^2)$
equation of CS and $q$-derivative of the $O(1)$ equation of
CS  (the moving-frame HF equation).
What if one adopts 
the $O(p^2)$ equation of CS or the  $q$-derivative of
the $O(1)$ equation of CS instead of the moving-frame RPA equation of
$O(p^2)$ ?
In the ASCC method without the higher-order operators,
three equations of motion are necessary, two of which are
the moving-frame HF equation and moving-frame RPA equation of $O(p)$.
To elucidate the role of $O(p^2)$ terms in the equation of CS,
below we investigate three cases
where, as the last one of the equations of motion, we adopt
(i) the $O(p^2)$ equation of CS,
(ii) the conventional moving-frame RPA equation of $O(p^2)$,
or 
(iii) the $q$-derivative of the $O(1)$ equation of CS.

\subsection{The $O(p^2)$ equation of collective submanifold}

When $\hat Q^{(2)}$ and $\hat Q^{(3)}$ are neglected,
the $O(p^2)$ equation of CS (\ref{eq:p^2 EoCSM}) reads
\begin{align}
\delta \langle \phi(q)|\frac{1}{2}[[\hat H,\hat Q^{(1)}],\hat Q^{(1)}] 
-B\partial_q\hat Q^{(1)} +\frac{1}{2}\partial_q B \hat Q^{(1)} | \phi(q) \rangle =0.
\end{align}
In the case of the Lipkin model, 
using Eqs. (\ref{eq:H in terms of tilde J0}), (\ref{eq:Q1 in terms of J}), 
and (\ref{eq:dqQ1 in terms of J}),
we obtain
\begin{align}
-\epsilon \chi\cos\phi\sin\phi \left( Q^{(1)}\right)^2
-B\partial_q Q^{(1)} +\frac{1}{2}\partial_q B Q^{(1)} =0.
\end{align}
Let us move to the coordinate system with $B\equiv 1$.
Using Eq. (\ref{eq:dqQ1}) and multiplying
both sides by $Q^{(1)}$, we find 
\begin{align}
\epsilon \chi\cos\phi\sin\phi \left( Q^{(1)}\right)^3
+\frac{1}{N} \partial_\phi Q^{(1)} =0
\,\,
\Leftrightarrow \,\,
-\frac{1}{\left( Q^{(1)}\right)^3}\partial_\phi Q^{(1)} 
=N\epsilon \chi\cos\phi\sin\phi .
\end{align}
Here we have used the canonical-variable condition
(\ref{eq:normalization of Q_A,P}).
The above equation is easily integrated 
%
with the initial condition (RPA solution) 
at $\phi=0$ shown in Eq. (\ref{eq: RPA solution at 0, pi}) as
\begin{align}
 Q^{(1)} 
=\frac{1}{\sqrt{2N}}  \left[ \epsilon +\epsilon\chi \left( 1- \frac{1}{2}\sin^2\phi\right) \right]^{-\frac{1}{2}}.
\end{align}
Obviously, this is inconsistent with $Q^{(1)}$ which is determined from
the $O(p)$ equation of CS (\ref{eq:Q1, P in ASCC without Q2}).
This inconsistency is thought to be caused by the neglect of  $\hat Q^{(2)}$ and $\hat Q^{(3)}$.
Thus, one cannot adopt the set of the $O(p)$ and $O(p^2)$ equations of
CS in this case.

\subsection{The moving-frame RPA equation of $O(p^2)$}
Next, we shall adopt the moving-frame RPA equation of $O(p^2)$.
We have already investigated this case in Sect. \ref{sec: soltion without Q2}.
In the case of the two-level Lipkin model,
$(Q^{(1)},P)$ are determined 
from the moving-frame RPA equation
of $O(p)$ and the canonical-variable condition only,
as shown in Eq. (\ref{eq:Q1, P in ASCC without Q2}).
Therefore,
there is no difference in the solution $(Q^{(1)},P)$ 
between the two cases where
the moving-frame RPA equation of $O(p^2)$ is adopted
and where the $q$-derivative of the $O(1)$ equation of CS is adopted. 
In both cases, $BC$ plays a role of the squared eigenfrequency of the
eigenvalue equations,
and it is in the eigenfrequency that there appears a difference between the two cases.
We have already seen the eigenfrequency squared for the moving-frame RPA equations
without the higher-order operators
in Sect. \ref{sec: soltion without Q2}, and it is given by Eq. (\ref{eq:BC in ASCC without Q2}).
%
If the set of the basic equations are self-consistent, the
eigenfrequency squared $\omega^2$ should
coincide
with the product of the inverse inertial mass $B$ and the potential curvature $C$.

We shall calculate the product of the inverse inertial mass $B$ and
the potential curvature $C$ in the conventional moving-frame HF \& RPA equations
without the higher-order operators.
The inverse inertial mass $B$ has been already obtained in Eq. (\ref{eq:B(phi) without Q2}).
The Christoffel symbol of the second kind is given by
\begin{align}
 \Gamma = -\frac{1}{2B}\frac{dB}{d\phi} =- \frac{1}{2P^2}\frac{d}{d\phi}P^2
= \frac{1}{2M}\frac{dM}{d\phi}= \frac{1}{2(Q^{(1)})^2}\frac{d}{d\phi}(Q^{(1)})^2
= \frac{1}{Q^{(1)}}\frac{dQ^{(1)}}{d\phi},
\end{align}
from which we have
\begin{align}
 \Gamma 
=- \frac{1}{2(2E+\epsilon\chi)}\frac{d}{d\phi}(2E+\epsilon\chi)
=\frac{1}{2(2E+\epsilon\chi)}\left(\epsilon\sin\phi-2\epsilon\chi\sin\phi\cos\phi \right).
\end{align}
Then, $C$ is calculated as
\begin{align}
 C&=\frac{d^2V}{d\phi^2}-\Gamma\frac{dV}{d\phi}\notag\\
&= \frac{1}{2}N \left(2E -\epsilon\chi\cos^2\phi\right)
-\frac{N}{4(2E+\epsilon\chi)}\epsilon^2\sin^2\phi\left( 1 -2\chi\cos\phi
 \right)
\left(1 -\chi\cos\phi \right).
\end{align}
Thus, we obtain
\begin{align}
BC=
\left(2E -\epsilon\chi\cos^2\phi \right)\left(2E+\epsilon\chi\right)
+\frac{1}{2}\epsilon^2\sin^2\phi\left(1 -\chi\cos\phi \right)
(2\chi\cos\phi-1) .
\label{eq:BC in phi-space}
\end{align}
The first term coincides with the squared eigenfrequency of the LRPA equations $\omega_{\rm LRPA}$,
in which the curvature term [the third term in Eq. (\ref{eq:moving-frame
RPA2 with Q2})] is omitted.
The second term is the contribution from the connection term, 
and the curvature term in the moving-frame RPA equations
should give this contribution.
As $BC$ is a scalar, the same result is obtained 
with the calculation in the $q$ space (See Appendix B).

Clearly, the squared eigenfrequency of the moving-frame RPA equation
without $\hat Q^{(2)}$
(\ref{eq:BC in ASCC without Q2})
and the product of the potential curvature and inertial mass
parameter $BC$ (\ref{eq:BC in phi-space}) are different by
\begin{align}
-\frac{1}{2}\epsilon^2\sin^2\phi(1-\chi\cos\phi)^2
=-2\left(\partial_qVQ^{(1)}\right)^2
=-\frac{2}{N^2}\left(\frac{dV}{d\phi}\right)^2,
\end{align}
and in this sense the self-consistency is broken.
This can be attributed to the fact that the higher-order operators are neglected
in the conventional moving-frame RPA equations.

\begin{figure}[tbp]
 \begin{center}
\includegraphics[width=0.5\textwidth]{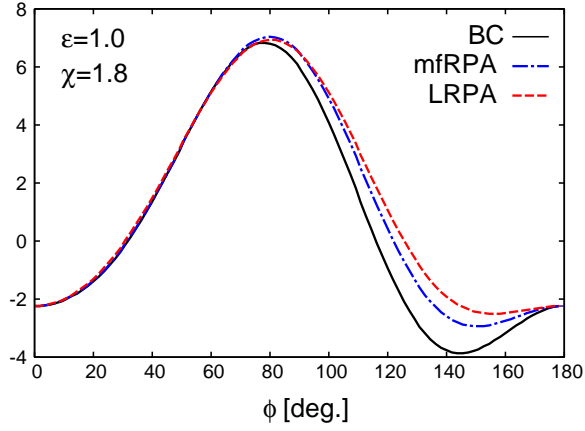}
 \end{center}
 \caption{%
The product of the potential curvature and inverse mass
$BC$ (solid) in comparison with the squared eigenvalues $\omega^2$ of the
 moving-frame RPA (mfRPA)
 equations without $Q^{(2)}$ (dot-dashed) and the LRPA equations
 (dashed) for $\chi=1.8, \epsilon=1.$
}
\label{BC_omega_mfLRPA_noQB}
\end{figure}

Fig. \ref{BC_omega_mfLRPA_noQB} shows the comparison of $BC$ with the
squared eigenfrequency of the
moving-frame RPA equations $\omega^2$ and that of the LRPA equations
$\omega^2_{\rm LRPA}$ without the higher-order operators.
One can see that the squared eigenfrequencies almost coincide with $BC$
for $\phi \lesssim \phi_0$, but their deviations from $BC$ become 
large for $\phi \gtrsim \phi_0$.

\subsection{The $q$-derivative of the $O(1)$ equation of collective submanifold}

Last, we consider the case where the $q$-derivative of the $O(1)$
equation of CS (\ref{eq:dq moving-frame HF eq.}) is adopted.
We shall obtain the eigenfrequency squared $BC=\omega^2$ in the eigenvalue problem given by the 
$O(p)$ equation of CS and the $q$-derivative of the $O(1)$ equation of
CS, and compare with the product of $B$ and $C$ we obtained in the
previous subsection.

By substituting Eqs.(\ref{eq:H in terms of tilde J0}), (\ref{eq:Q1 in terms of J}), 
(\ref{eq:BP in terms of J}), and (\ref{eq:dqQ1 in terms of J}) into
Eq. (\ref{eq:dq moving-frame HF eq.}),
we obtain
\begin{align}
\left[ \epsilon\cos\phi + \epsilon\chi(\sin^2\phi-\cos^2\phi) \right]BP
 -BCQ^{(1)} 
+\frac{1}{2}\partial_q B \partial_qV Q^{(1)}
-B\partial_q V \partial_q Q^{(1)} =0.
\label{eq:dqO(1) without Q2}
\end{align}

The third term vanishes when we employ the coordinate system with
$B\equiv 1$.
%
By rewriting $\partial_q V$ and $\partial_qQ^{(1)}$ 
with use of Eqs. (\ref{eq:mfHF with Lipkin})  and (\ref{eq:dqQ1}), respectively,  
$BC$ can be rewritten in terms of $(Q^{(1)}, P)$. 
Then, with Eq. (\ref{eq:Q1, P in ASCC without Q2}), we readily obtain
\begin{align}
BC&=\left[ \epsilon\cos\phi + \epsilon\chi(\sin^2\phi-\cos^2\phi) \right]
\left[ \epsilon\cos\phi + \epsilon\chi(1+\sin^2\phi) \right]\notag\\ 
& +\frac{1}{2}\epsilon^2\sin^2\phi(1-\chi\cos\phi) (2\chi\cos\phi-1),
\label{eq: omega2 in 1st-order ASCC with dq mfHF}
\end{align}
Unlike the conventional moving-frame RPA equations without $\hat Q^{(2)}$,
the eigenfrequency squared (\ref{eq: omega2 in 1st-order ASCC with dq mfHF}) coincides with $BC$ (\ref{eq:BC in phi-space}) 
obtained from the potential curvature and the inverse inertial mass,
so they are  self-consistent in this case.
While the higher-order operators are involved in the moving-frame RPA
equation of $O(p^2)$, which were neglected in the previous subsection, 
the $q$-derivative of the $O(1)$ equation of CS 
does not contain the higher-order operators, and no approximation is made.
In this sense, the $q$-derivative of the $O(1)$ equation of CS is better
than the moving-frame RPA equation of $O(p^2)$ with the higher-order
operators ignored. 
It is noteworthy that $\hat Q^{(2)}$ is omitted in the $O(p)$ equation
of CS in both of the two cases.



\section{Discussion}

So far we have considered a method of determining $\hat Q^{(2)}$
in the case without the pairing correlation.
As shown in Refs. \cite{Sato2017a, Sato2017b},
the higher-order operators have much to do with the gauge symmetry.
In this section, we briefly discuss the gauge symmetry
of the basic equations when the pairing correlation is included.

First we shall reconsider the gauge symmetry that
Hinohara et al found~\cite{Hinohara2007}.
As mentioned in Introduction, Hinohara et al 
encountered the numerical instability
caused by the gauge symmetry of the basic equations,
and needed to introduce a gauge-fixing prescription
for successful calculation.
This is a ''numerical'' problem in the sense explained below.
The moving-frame HFB \& QRPA equations including up to  the first-order
operator $\hat Q^{(1)}$
are given by,
\begin{equation}
 \delta \langle \phi(q)|\hat H_M
  |\phi(q)\rangle =0,
\label{eq:moving-frame HFM}
\end{equation}

\begin{equation}
 \delta \langle \phi(q)|[\hat H_M , \hat Q^{(1)}] 
-\frac{1}{i}B(q)\hat P |\phi(q)\rangle =0,
\label{eq:moving-frame QRPA1 without Q2}
\end{equation}

\begin{align}
& \delta \langle \phi(q)|
[\hat H_M, \frac{1}{i}B\hat P]
-B(q)C(q)\hat Q^{(1)}-\partial_q\lambda \tilde N 
-\frac{1}{2}
\partial_q V
[[\hat H_M,  \hat Q^{(1)}],\hat Q^{(1)}] 
|\phi(q)\rangle =0. \label{eq:moving-frame QRPA2 without Q2}
\end{align}
with $\hat H_M =\hat H -\lambda \tilde N -\partial_qV \hat Q^{(1)}$.
(Assume that $\hat Q^{(1)}$ contains only A-terms.)
When $\partial_qV=0$, these equations are invariant under the
transformation (\ref{eq:^Q' ex1})-(\ref{eq:dqlambda' ex1}).
When $\partial_qV \neq 0$, the basic equations are not invariant
under this transformation. Nevertheless,
there occurred the numerical instability.
This may be understood as follows.

As shown in Ref. \cite{Sato2017a}, for $\partial_qV\neq 0$, 
the moving-frame HFB equation (\ref{eq:moving-frame HFM}) is invariant under this gauge transformation, but
the moving-frame QRPA equations (\ref{eq:moving-frame QRPA1 without Q2}) 
and (\ref{eq:moving-frame QRPA2 without Q2})  are not. 
Under the transformation, $[\hat H_M, \hat Q^{(1)}]$ transforms as below,
\begin{align}
 [\hat H_M, \hat Q^{(1)}] \rightarrow [\hat H_M, \hat Q^{(1)}] -\alpha
 \partial_qV[\tilde N, \hat Q^{(1)}],
\end{align}
so the $O(p)$ moving-frame QRPA equation (\ref{eq:moving-frame QRPA1 without Q2}) is not gauge invariant.
The second term of the right-hand side comes from $\partial_q V \hat Q^{(1)}$ in $\hat H_M$.
As for the moving-frame QRPA equation of 
$O(p^2)$ (\ref{eq:moving-frame QRPA2 without Q2}), the curvature term
\begin{align}
 -\frac{1}{2}\partial_q V[[\hat H_M,  \hat Q^{(1)}],\hat Q^{(1)}]
\end{align}
breaks the gauge symmetry as easily confirmed.
Hinohara et al started the calculation from the HFB 
equilibrium point ($q=0$) where $\partial_q V(q=0)=0$,
and at the next step $q=\pm \delta q$, $\partial_q V(\delta q)$ is still small.
Although there are gauge-symmetry-breaking terms in the moving-frame
QRPA equations as seen above, 
they are proportional to $\partial_q V$, and their contributions are
small in the vicinity of the HFB equilibrium point $\partial_q V(q)=0$.
Thus, the gauge symmetry is approximately retained, 
which leads to the numerical instability.
If the gauge-symmetry-breaking terms gave sufficiently large
contributions, the numerical instability would not occur. 
Then, once the gauge is fixed at the HFB equilibrium point,
the gauge-fixing prescription would not be necessary to solve the
moving-frame equations
at non-equilibrium points.
The term $[[H_M,\hat Q^{(1)}],\hat Q^{(1)}]$ originates from
the $O(p^2)$ equation of CS [See Eq. (\ref{eq:p^2 EoCSM with pairing})]
and breaks the gauge symmetry (in other
words, fixes the gauge).
It is multiplied by $\partial_q V$ 
to remove $D_q Q^{(1)}$ from the $q$-derivative of the $O(1)$ equation
of CS in the derivation of the moving-frame QRPA equation of $O(p^2)$,
which leads to the small contribution in the moving-frame QRPA equation.

Next, we investigate the gauge symmetry of the basic equations
we have proposed in this paper.
First, let us assume the case where only the operators up to the first order 
are taken, and  Eq. (\ref{eq:dq moving-frame HFB eq.}) is
adopted instead of the moving-frame QRPA equation of $O(p^2)$.
Concerning the gauge symmetry of Eq.
(\ref{eq:dq moving-frame HFB eq.}), the part
\begin{align}
 [\hat H_M, \frac{1}{i}B(q)\hat P]
-B(q)C(q)\hat Q^{(1)}-B\partial_q\lambda \tilde N
\end{align}
is gauge invariant, then what we have to check is the
gauge symmetry of $\partial_qV D_q \hat Q^{(1)}$.
It is sufficient to  investigate the gauge symmetry
in the vicinity of the HFB equilibrium point $\partial_q V=0$.
We choose the coordinate system with $B(q)=1$, and then $D_q \hat
Q^{(1)}=\partial_q \hat Q^{(1)}$.
In the previous section, due to the simplicity of the Lipkin model, the basic equations
reduced to one differential equation.
In general cases, however, $\partial_q \hat Q^{(1)}$ should be approximated by
finite difference. With the simplest scheme,
\begin{align}
  B\partial_q V(\delta q) \partial_q \hat Q(\delta q) 
=& B\partial_q V(\delta q) \frac{1}{\delta q} \left[\hat Q^{(1)}(\delta q) -\hat Q^{(1)}(0) \right]
\notag \\
=& B\left[ \partial_q V(0) + \partial_q^2 V(0) \delta q  \right]
\frac{1}{\delta q} \left[\hat Q^{(1)}(\delta q) -\hat Q^{(1)}(0) \right]
\notag \\
=& BC(0) \hat Q^{(1)}(\delta q) - BC(0)\hat Q^{(1)}(0) \notag \\
\rightarrow & BC(0) \hat Q^{(1)}(\delta q) - BC(0)\hat Q^{(1)}(0) +\alpha
 BC(0)\tilde N.
\end{align}
It is not gauge invariant except for the zero modes.

Then, we investigate the gauge symmetry in the case where
we adopt Eq. (\ref{eq:moving-frame HFB})-(\ref{eq:p^2 EoCSM with
pairing}) as the basic equations.
As shown in Ref. \cite{Sato2017a}, Eqs. (\ref{eq:moving-frame
HFB})-(\ref{eq:moving-frame QRPA1 with Q2}) are gauge invariant.
Eq. (\ref{eq:dq moving-frame HFB eq.}) does not contain $\hat Q^{(2)}$,
so the consideration above holds as it is.
Unless $C(0)=0$, 
the gauge symmetry is significantly broken even in the vicinity of HFB
equilibrium point $\partial_q V=0$, and the gauge is fixed.
The term $[\hat Q^{(1)},\hat Q^{(2)}]$ in Eq. (\ref{eq:p^2 EoCSM with
pairing})
is a B-term and does not contribute, so 
Eq. (\ref{eq:p^2 EoCSM with pairing})  with this term omitted, 
\begin{align}
 \delta \langle \phi(q)|\frac{1}{2}[[\hat H_M ,\hat Q^{(1)}], \hat Q^{(1)}] 
-\partial_q\hat Q^{(1)} 
-\frac{i}{2}[\hat H-\lambda \tilde N ,\hat Q^{(2)}]  
|\phi(q)\rangle =0,  
\end{align}
would be solved in the actual
calculation. 
It is worth mentioning that
the moving-frame QRPA equation of $O(p^2)$ and the $O(p^2)$ equation of
CS with the $\hat Q^{(3)}$ term, which is ignored here,  are gauge-invariant if
the $[\hat Q^{(1)}, \hat Q^{(2)}]$ term is included (See Ref. \cite{Sato2017a}).

Under the gauge transformation, the first and third terms $(\times 2)$ transform as
\begin{align}
 &[[H_M,\hat Q^{(1)}],\hat Q^{(1)}]-i[\hat H-\lambda \tilde N, \hat
 Q^{(2)}] \notag\\
\rightarrow  &[[H_M,\hat Q^{(1)}],\hat Q^{(1)}]-i[\hat H-\lambda \tilde
 N, \hat  Q^{(2)}]
+\alpha \partial_qV 
\left(
[[\tilde N, \hat Q^{(1)}],\hat Q^{(1)}] 
-i[\tilde N,\hat Q^{(2)}]\right)
+ O(\alpha^2
 \partial_qV),
\end{align}
so the gauge-symmetry-breaking term is proportional to $\partial_q V$.
The rest transforms as
\begin{align}
 \partial_q \hat Q^{(1)}(\delta q) 
=\left[\hat Q^{(1)}(\delta q) -\hat Q^{(1)}(0)\right]/\delta q 
\rightarrow   \partial_q \hat Q^{(1)}(\delta q) +\frac{\alpha}{\delta
 q}\tilde N.
\end{align}
This term breaks the gauge symmetry, and the gauge can be regarded as fixed.


\section{Concluding remarks}
In this paper, we have considered the ASCC theory 
including the second-order collective-coordinate operator $\hat Q^{(2)}$
which consists of only A-terms,  proposed a new set of basic
equations to determine the collective operators including $\hat Q^{(2)}$,
and applied it to the two-level Lipkin model.
We have compared the ASCC calculations with and without $\hat Q^{(2)}$
and found that, for a first few low-energy states, the difference between the
results of the two calculations is not so significant and that
both of the calculations reproduce the exact solution well. 
However, with increasing the excitation energy, 
the deviation from the exact solution becomes larger
in the case without $\hat Q^{(2)}$, while, with $\hat Q^{(2)}$, the
agreement with the exact solution is good even for the higher excited states.
As discussed in Refs. \cite{Sato2017a, Sato2017b} and this paper,
$\hat Q^{(2)}$ does contribute to the inertial mass.
As the excitation energy increases, the kinetic energy becomes
important relatively to the collective potential energy.
The results shown in this paper illustrates the importance of the
correct evaluation of the inertial mass.

We have also reconsidered the basic equations to adopt 
in the case where no higher-order operator is included.
It has been shown that, in the conventional moving-frame RPA equations,
the self-consistency is broken in the sense that
the eigenfrequency squared $\omega^2$ does not coincide with
the product of the potential curvature and the inverse inertial mass $BC$.
In contrast,
when we employ the $q$-derivative of the $O(1)$ equation of CS,
in which no approximation is made for the higher-order operators,
the relation $\omega^2=BC$ holds, and the self-consistency is kept.
In the case of the two-level Lipkin model we have used,
$(\hat Q^{(1)},\hat P)$ are determined from the $O(p)$ equation of CS
and the canonical-variable condition, and the  difference between 
the moving-frame RPA equation of $O(p^2)$ and 
the $q$-derivative of the $O(1)$ equation of CS
appears only in $\omega^2$.
It would be interesting to investigate 
how $(\hat Q^{(1)},\hat P)$ are affected depending on which of the two equations to
adopt, using the three-level Lipkin or more realistic models.
The observation above may lead to an intuitive understanding as follows.
The $O(1)$ equation of CS is an equation for the state vector
$|\phi(q)\rangle$ and the Lagrange multiplier, 
once $\hat Q^{(1)}$ is given.
The $O(p)$ equation of CS and the $q$-derivative of the $O(1)$ equation
of CS can be viewed as equations to determine $\hat Q^{(1)}$ and $\hat P$, respectively,
while $\hat Q^{(2)}$ is determined from the $O(p^2)$ equation of CS.
Note that all the basic equations should be solved self-consistently.

Although we have mainly studied the ASCC theory without the pairing correlation in this paper,
the basic equations with the pairing correlation
are also derived in a straightforward way,
and we have briefly discussed the gauge symmetry of the basic equations.
The gauge transformation changes the gauge angle and the chemical
potential as well as the collective operators, and it plays
an important role in the treatment of superfluid systems.

As shown in Refs. \cite{Sato2015, Sato2017a}, the equation of CS 
before the adiabatic expansion is gauge invariant, but the gauge
symmetry is (partially) broken by the adiabatic expansion. 
In Refs. \cite{Sato2015, Sato2017a}, 
we analyzed the gauge symmetry under the general gauge transformation,
and found that four examples or types of the gauge transformations 
play an essential role in the analysis.
One of the four is the gauge transformation Eqs.(\ref{eq:^Q' ex1})
-(\ref{eq:dqlambda' ex1}) 
(We call it Example 1 in Refs. \cite{Sato2015, Sato2017a}).  
The gauge symmetry of Example 1 can be retained 
by including the higher-order collective-coordinate 
operators. However,
the symmetry under the gauge transformation of Example 3 in 
Ref. \cite{Sato2015, Sato2017a}, 
which mixes $\hat P$ with $\tilde N$, cannot be retained 
even if the higher-order operators are introduced.
This can be regarded as a gauge fixing by the adiabatic expansion.
In Hinohara's prescription, they require 
only the gauge symmetry under the transformation which mixes $\hat Q^{(1)}$
with $\tilde N$ (Example 1),  
and the gauge symmetry under the transformation which mixes 
$\hat P$ with $\tilde N$ (Example 3) is left broken.
In this sense, Hinohara's prescription 
attaches more weight to Example 1 than Example 3.
If the two transformations are of equal weight, the basic equations 
breaking the symmetry under the gauge transformation of Example 1 
can be adopted.
Then one may regarded it as a gauge fixing as in Example 3.
In general, if there is gauge symmetry, one can choose a convenient
gauge, so a set of the basic equations which are not gauge invariant 
may be possible.

As mentioned in Introduction, the extension of the ATDHF to the ATDHFB
theory is not straightforward, because one has to decouple the
number-fluctuating mode from the collective mode of interest.
For the collision of two nuclei, the chemical potentials
of the two nuclei are different unless it is a collision between the same nuclides.
To describe such a phenomenon, it is necessary to construct a theory 
with which one can treat the gauge degrees of freedom correctly.
In Ref. \cite{Wen2017},
Wen and Nakatsukasa could not find
a collective path connecting the superdeformed state and the ground state
in $^{32}$S, and pointed out a possible improvement of the problem
by including the pairing correlation.
For that purpose, a theory is required which can treat both cases with and
without the pairing correlation on an equal footing.
The basic equations we have proposed in this paper has such an advantage
and treat the higher-order collective operator in both of the cases on
an equal footing.
It would be very interesting to apply the set of the basic equations
proposed here to systems with the pairing correlation,
and it will be reported in a future publication.

\section*{Acknowledgements}
The author thanks K. Matsuyanagi, M. Matsuo, T. Nakatsukasa, and N. Hinohara for fruitful comments.

\appendix

\section{Derivatives of the collective operators}

We shall obtain the derivative of the operator $\hat O(q)$
which is written in terms of the quasispin operators by evaluating
\begin{align}
\partial_q \hat O(q)&= \lim_{\delta q \rightarrow 0} \frac{\hat O(q)
 -\hat O(q-\delta q)}{\delta q}.
\end{align}
Here we consider the following three types of operators
corresponding to $\hat Q^{(1)},\hat P$ and $\hat Q_B$.
\begin{align}
 \hat E_{A}(q-\delta q)&= E_{A}(q-\delta q)(\hat J_+(q-\delta q)+\hat J_-(q-\delta q)),\\
 \hat O_{A}(q-\delta q)&=iO_{A}(q-\delta q)(\hat J_+(q-\delta q)-\hat J_-(q-\delta q)),\\
 \hat E_{B}(q-\delta q)&= E_{B}\tilde J_0(q-\delta q)
= E_{B}\left(\hat J_0(q-\delta q)+\frac{N}{2}\right).
\end{align}
While $\hat Q_B=Q_B\tilde J_0$ is not necessary in this paper, 
we have added it for completeness.

To evaluate the derivative, one has to rewrite 
the quasispin operators at $q-\delta q$,
$\hat J_\pm(q -\delta q)$ and $\hat J_0(q-\delta q)$ 
in terms of $\hat J_\pm(q)$ and $\hat J_0(q)$ at $q$.
Let us denote Eq. (\ref{eq:Hol(16)}) by
\begin{align}
 \begin{pmatrix}
  \hat I_0(q)\\ \hat I_+(q) \\ \hat I_-(q) 
 \end{pmatrix}
=
A(\phi(q),\psi(q))
 \begin{pmatrix}
  \hat K_0\\ \hat K_+ \\ \hat K_- 
 \end{pmatrix}.
\end{align}
We define  $A(\phi(q)) :=A(\phi(q),\psi(q)=0)$,
and then
\begin{align}
 \begin{pmatrix}
  \hat J_0(q-\delta q)\\ \hat J_+ (q-\delta q)\\ \hat J_-(q-\delta q) 
 \end{pmatrix}
&=
A(\phi(q-\delta q))
 \begin{pmatrix}
  \hat K_0\\ \hat K_+ \\ \hat K_- 
 \end{pmatrix}
=
A(\phi(q-\delta q))A(\phi(q))^{-1}
 \begin{pmatrix}
  \hat J_0(q)\\ \hat J_+ (q)\\ \hat J_-(q) 
 \end{pmatrix}.
\end{align}
Let us simplify the notation with $(\phi(q-\delta q),\phi(q))=(\phi_o,\phi_n$), 
and then 
\begin{align}
&A(\phi(q-\delta q))A(\phi(q))^{-1}=A(\phi_o)A(\phi_n)^{-1}
\notag\\
=&
 \begin{pmatrix}
  \cos\phi_o && -\frac{1}{2} \sin\phi_o &&  -\frac{1}{2}\sin\phi_o \\ 
  \sin\phi_o &&  \cos^2\frac{1}{2}\phi_o && -\sin^2\frac{1}{2}\phi_o \\ 
  \sin\phi_o && -\sin^2\frac{1}{2}\phi_o &&   \cos^2\frac{1}{2}\phi_o 
 \end{pmatrix}
\begin{pmatrix}
\cos \phi_n && \frac{1}{2}\sin\phi_n && \frac{1}{2}\sin \phi_n \\
-\sin\phi_n && \cos^2 \frac{1}{2}\phi_n && -\sin^2 \frac{1}{2}\phi_n \\
-\sin \phi_n && -\sin^2\frac{1}{2}\phi_n && \cos^2 \frac{1}{2}\phi_n
\end{pmatrix}\notag\\
=&
 \begin{pmatrix}
  \cos(\phi_o-\phi_n) && -\frac{1}{2} \sin(\phi_o-\phi_n) &&  -\frac{1}{2}\sin(\phi_o-\phi_n) \\ 
  \sin(\phi_o-\phi_n) &&  \cos^2\frac{1}{2}(\phi_o-\phi_n) && -\sin^2\frac{1}{2}(\phi_o-\phi_n) \\ 
  \sin(\phi_o-\phi_n) && -\sin^2\frac{1}{2}(\phi_o-\phi_n) &&   \cos^2\frac{1}{2}(\phi_o-\phi_n) 
 \end{pmatrix}.
\label{eq:transform J}
\end{align}

Now the operators at $q-\delta q$ are readily rewritten in terms of the quasispins
at $q$.
\begin{align}
\hat E_{A}(q-\delta q)
&= \left( E_{A}(q)-\delta q\partial_q E_{A}  \right)
\left[
-2\sin \delta \phi \hat J_0(q)+
\cos\delta\phi \left( \hat J_+(q)+\hat J_-(q) \right)\right],
\end{align}
where $\delta\phi=\phi_n-\phi_o=\phi(q)-\phi(q-\delta q)$, 
which leads to
\begin{align}
&\frac{\left(\hat E_{A}(q) - \hat E_{A}(q-\delta q)\right)}{\delta q}\notag\\
=& 
\frac{2\sin \delta \phi}{\delta \phi}\frac{\delta \phi}{\delta q}E_A(q)\hat J_0(q)
+\partial_q E_{A}  
\cos\delta\phi \left( \hat J_+(q)+\hat J_-(q) \right) + O(\delta q).
\end{align}
Taking the limit $\delta q\rightarrow 0$, we have
\begin{align}
 \partial_q \hat E_A
=2E_A(q)\frac{\partial\phi}{\partial q}\hat J_0(q)
+\partial_q E_{A} \left( \hat J_+(q)+\hat J_-(q) \right) .
\end{align}
The derivative of a time-even operator of A-type $\hat E_A$ contains both of A-terms
and B-terms. 
Similarly,
\begin{align}
\hat O_{A}(q-\delta q)
&= i\left( O_{A}(q)-\delta q\partial_q O_{A}  \right)
\left( \hat J_+(q)-\hat J_-(q) \right),
\end{align}
from which we obtain
\begin{align}
 \partial_q \hat O_A(q)
=i\partial_q O_{A} \left( \hat J_+(q)-\hat J_-(q) \right) .
\end{align}
The derivative of the B-type operator can be calculated similarly.
\begin{align}
 &\hat E_{B}(q-\delta q) \notag\\
=& \left( E_{B}(q)-\delta q\partial_q E_{B}  \right)
\left[
\cos\delta\phi\left( \hat J_{0}(q)+\frac{N}{2}\right)
+\frac{1}{2}\sin\delta\phi
\left( 
\hat  J_+(q)+\hat J_-(q) \right)
\right],
\end{align}
from which we have
\begin{align}
 \partial_q \hat E_B
=
\partial_q E_{B} \tilde J_0(q)
-
\frac{1}{2}
E_B(q)\frac{\partial\phi}{\partial q}
\left( \hat J_+(q)+\hat J_-(q) \right) .
\end{align}
The derivatives of the collective operators now read
\begin{align}
 \partial_q\hat Q^{(1)}(q)
&=
\partial_q Q^{(1)}(q) \left( \hat J_+(q)+\hat J_-(q) \right) 
+\frac{2}{N}\hat J_0(q),\\
 \partial_q \hat P(q)&=i\partial_q P(q) \left( \hat J_+(q)-\hat J_-(q) \right) ,\\
 \partial_q\hat Q_B(q)
&=
\partial_q Q_{B}(q) \tilde J_0(q)
-Q_B(q)P(q)
\left( \hat J_+(q)+\hat J_-(q) \right). 
\end{align}
Here we have used Eq. (\ref{eq:dphi dq}). 
The second terms of $\partial_q \hat Q^{(1)}$ and $\partial_q \hat Q_B(q)$
coincide with $\frac{1}{i}[\hat P,\hat Q^{(1)}]$ and $\frac{1}{i}[\hat P,\hat Q_{B}]$, respectively.

\section{Curvature in the $q$ space}
In this Appendix, we shall evaluate the potential curvature $C(q)$ in the
$q$ space when only the first-order collective-coordinate operator is
taken into account.
The collective potential $V(\phi)$ as a function of $\phi$ is given by
Eq. (\ref{eq:HF energy}).

In the  $q$ space with $B(q) \equiv 1$,
the curvature $C(q)$ is given by
\begin{align}
C(q)=\frac{d^2V}{dq^2}
=\frac{d}{dq}\left(\frac{dV}{d\phi}\frac{d\phi}{dq}\right)
 =\frac{d^2V}{d\phi^2}\left(\frac{d\phi}{dq}\right)^2+\frac{dV}{d\phi}\left(\frac{d^2\phi}{dq^2}\right).
\label{app eq:dV^2/dq^2}
\end{align}
From Eqs. (\ref{eq:dphi dq}) and (\ref{eq:Q1, P in ASCC without Q2}),
the first term of Eq. (\ref{app eq:dV^2/dq^2}) reads
\begin{align}
 \frac{d^2V}{d\phi^2}\left(\frac{d\phi}{dq}\right)^2
&=\frac{1}{2}N \left( 2E -\epsilon\chi\cos^2\phi \right)\cdot 4P^2
=\left(2E -\epsilon\chi\cos^2\phi
 \right)\left(2E+\epsilon\chi\right)=\omega_{\rm LRPA}^2.
\end{align}
$\omega_{\rm LRPA}$ is the eigenfrequency of the LRPA equations,
in which the curvature term is neglected.
Similarly, from Eqs. (\ref{eq:dphi dq}) and (\ref{eq:Q1, P in ASCC
without Q2}), we have
\begin{align}
 \frac{d^2\phi}{dq^2}&=2\frac{dP}{dq}
=2\frac{dP}{d\phi}\frac{d\phi}{dq}
=4P\frac{dP}{d\phi}
=\frac{1}{N}\epsilon\sin\phi(2\chi\cos\phi-1).
\end{align}
Then, the  second term of Eq. (\ref{app eq:dV^2/dq^2}) is 
\begin{align}
\frac{dV}{d\phi}\left(\frac{d^2\phi}{dq^2}\right)
&=\frac{1}{2}\epsilon^2\sin^2\phi\left(1 -\chi\cos\phi \right)
(2\chi\cos\phi-1).
\end{align}
Thus, the product of the curvature and the inverse mass [$B(q) \equiv 1$] is obtained as
\begin{align}
BC=C(q)
=
\left(2E -\epsilon\chi\cos^2\phi \right)\left(2E+\epsilon\chi\right)
+\frac{1}{2}\epsilon^2\sin^2\phi\left(1 -\chi\cos\phi \right)
(2\chi\cos\phi-1),
\label{eq:C in q-space}
\end{align}
which coincides with Eqs. (\ref{eq:BC in phi-space})  
and (\ref{eq: omega2 in 1st-order ASCC with dq mfHF}).
The second term vanishes when $2\chi\cos\phi-1=0$ as well as at the potential extrema.
One can immediately find that $B(\phi)C(\phi)=B(q)C(q)$ by noticing that
from Eqs. (\ref{eq:dphi dq}) and  (\ref{eq:M(phi) in terms of Q1 and P}),
\begin{align}
 \left(\frac{d\phi}{dq}\right)^2=B(\phi), \,\,\,
 \frac{d^2\phi}{dq^2}=4P\frac{dP}{d\phi}=\frac{1}{2}\frac{dB(\phi)}{d\phi}=-\Gamma(\phi)B(\phi).
\end{align}


\end{document}